\documentclass{aa}  

\usepackage{graphicx}
\usepackage{txfonts}
\usepackage{multicol}
\usepackage{xcolor}
\begin{document}

   \title{LISA test-mass charging. Particle flux modeling, Monte Carlo simulations and  induced effects on the sensitivity of the observatory}

   \subtitle{}
   \author{F. Dimiccoli\inst{1,2}, R. Dolesi\inst{1,2}, M. Fabi\inst{3,4}, V. Ferroni\inst{1,2}, C. Grimani\inst{3,4}, M. Muratore\inst{5}, P. Sarra\inst{6} M. Villani\inst{3,4} and W.J. Weber\inst{1,2}
          }
    \institute{Trento Institute for Fundamental Physics and Applications (TIFPA-INFN), Trento, Italy \and Department of Physics, University of Trento, Trento, Italy\\
             \email{francesco.dimiccoli@unitn.it}
             \thanks{corresponding author}
            \and University of Urbino Carlo Bo, Department of Pure and Applied Sciences (DiSPeA), Via Santa Chiara, 27, Urbino (PU), 61029, Italy
            \and National Institute for Nuclear Physics (INFN), Section in Florence, Via B. Rossi 1, Florence, 50019, Italy
            \and Max Planck Institute for Gravitational Physics (Albert Einstein Institute), D-14476 Potsdam, Germany
            \and OHB Italia SpA Milano, Italy}

   \date{}

   \abstract
   {The LISA space observatory will explore the sub-Hz spectrum of  gravitational wave emission from the Universe. The space environment, where will be immersed in, is responsible for charge accumulation on its free falling test masses (TMs) due to the galactic cosmic rays (GCRs) and solar energetic particles (SEP) impinging on the spacecraft. Primary and secondary particles produced in the spacecraft material  eventually reach the TMs by depositing a net positive charge fluctuating in time. This work is relevant for any present and future space missions that, like LISA, host free-falling TMs as inertial reference.}
   {The coupling of the TM charge with native stray electrostatic field produces  noise forces on the TMs, which can limit the performance of the LISA mission. A precise knowledge of the charging process allows us to predict the intensity of these charge-induced disturbances and to design specific counter-measures.}
   {We present a comprehensive toolkit that allows us to calculate the TM charging time-series in a geometry representative of LISA mission, and the associated induced forces under different conditions of the space environment by considering the effects of short, long GCR flux modulations and  SEPs.}
   {We study, for each of the previously mentioned conditions, the impact of spurious forces associated with the TM charging process on the mission sensitivity for gravitational wave detection.}
   {}


   \maketitle

\section{Introduction: an end-to-end toolkit for the simulation of the LISA test-mass charging  \label{introduction}}

Experimental gravitational wave physics had a turning point in 2015 when the LIGO interferometers detected the gravitational radiation emission from the merging of a stellar black hole binary system  \citep{gw_discovery}. That was the first direct confirmation of the existence of gravitational waves predicted by Einstein's theory of General Relativity \citep{einstein16}. 

The ESA LISA Pathfinder (LPF) mission \citep{lisapf1,antonucci2011} was launched on December 4th of the same year \citep{armano2016,armano2017a}. This mission demonstrated that a reference mass (same as test-masses, TMs, in the following) can be put into geodesic motion with a residual stray acceleration of the level required by a space interferometer for gravitational wave observation built in space. The Laser Interferometer Space Antenna (LISA) has been approved by ESA in its phase A in 2017 \citep{amaro2017}, with the aim to open a window on the mHz band \citep{lisa_redbook}, and finally adopted in January 2024, with expected launch in 2035.  LISA will use free-falling test masses as end mirrors of a large space interferometer with 2.5 million km arms. The LISA TMs will be hosted on a constellation of three spacecraft (S/C) in triangular formation in orbit around the Sun, trailing Earth at about 50 million km. 
The LISA  sensitivity is expected to unveil the gravitational wave spectrum populated by a large sample of galactic binaries, massive black holes, extreme-mass ratio inspirals  and more exotic sources \citep{lisa_redbook}. LISA-based observations will be complementary to those gathered by Earth interferometers, being capable of detecting gravitational waves when the binary sources arrive at
merging in their measurement frequency band above a few Hz,sometimes
after having emitted in the inspiral phase in the LISA band (multiband astronomy, \cite{PhysRevLett.116.231102}). Moreover, LISA will be synergic with other missions observing in the electromagnetic band \citep{lisagrb} and Earth neutrino experiments  for a multimessenger approach to the physics of compact objects \citep{suwa}.
 
 The LISA TM has to be in free-fall to within a stray acceleration noise below 3 fm s$^{-2}$ Hz$^{-1/2}$ at 1 mHz. The space environment eliminates stray forces from Newtonian gravitational noise, seismic noise, and suspension thermal noise, which are the key limiting factors on ground. However, it affects differently the free-fall status of the TMs. The LISA's TMs are 46 mm gold-platinum alloy cubes coated with a 800 nm thick gold layer. The TMs are surrounded, with no mechanical contact and few mm gap, by  a gold-plated electrode-housing (EH) used for both position sensing and actuation \citep{lisapf2}. High-energy particles of galactic and solar origin charge the LISA TMs making them sensitive to Coulomb forces due to stray electric fields  found between the masses and the surrounding EH \citep[see for details][]{lisapf2}. 
 These Coulomb forces  a) contribute to the acceleration noise of the TMs due to the intrinsic time fluctuations of the deposited charge and of the stray electric fields and b) generate spurious signals in the LISA sensitivity band due to solar energetic particle (SEP) events and Forbush decreases (FDs) \citep{forbush1,forbush2,forbush3,apj1,apj2,apj3}, sudden depressions of the galactic cosmic-ray (GCR) flux due to the passage of interplanetary counterparts of coronal mass ejections (ICMEs). 
 
 The  charging process is associated with energetic particles able to penetrate about $16\,\mathrm{g\ cm}^{-2}$ of shielding material surrounding the TMs \citep{geomlisa} expected to be provided by the LISA payload and spacecraft.
 This corresponds to a minimum energy threshold  of approximately $100\,\mathrm{MeV}$, $20\,\mathrm{MeV}$ and $100\,\mathrm{keV}$ respectively for  hadrons, electrons and photons of galactic and solar origin penetrating or interacting in the spacecraft \citep{lisagrb}.  
 The characterization of the charging process is primarily linked to the evaluation of two fundamental rate parameters (see proper definition in Eqn. \ref{lnet} and \ref{leff}, Section \ref{studysection}): $\lambda_{NET}$, indicating the rate at which the TM accumulates charge, and $\lambda_{EFF}$, which is the effective single charge Poissonian event rate associated with TM charge noise \citep{araujo}.

Monte Carlo simulations of the LPF TM charging were carried out with FLUKA \citep{flukacern1,flukacern2,flair} and GEANT4 \citep{GEANT4:2002zbu,Allison:2016lfl,1610988} before the mission was launched at the end of 2015. These simulations returned very similar results for both TM charging rate and noise at solar minimum conditions \citep{araujo,grim15,wass2005}, despite different nominal energy thresholds for electron propagation were considered in GEANT4 ($250\,\mathrm{eV}$) and FLUKA ($1\,\mathrm{ keV}$). This evidence was initially explained only in terms of the average ionization potential in gold that limited the hadron ionization to 790 eV in GEANT4 \citep{mattia,taioli23}. Conversely, in this work we show that the main problem comes from the generation and propagation of very low energy electrons.

The mission pre-launch work in \cite{grim15} was carried out on the basis of predictions of the solar modulation between the end of 2015 and beginning 2016. The results of Monte Carlo simulations carried out with these updated predictions were compared to the LPF TM charging measurements  carried out in space in April 20-23, 2016
\citep{Armano2017} (see Table \ref{tab:LPF_Results}). The measured net charging appeared to lie in the middle of the prediction range, while the effective charging was three to four times higher than expected. Moreover, it was observed that the net charging depended on the potential of the TM (V$_{TM}$), saturating to 0 for V$_{TM} \approx$ 1 V \citep{armanoTM23}. 
This observational scenario was ascribable to particles with the same charge sign entering and escaping the TMs in approximately equal number, thus contributing to the charging noise but not to the net charging. Low-energy electrons (LEE, with E<1 keV ) produced at the gold surfaces of the TMs and EH were the most plausible candidates 
to explain the observations as originally suggested  in  \citet{araujo}.  In fact Monte Carlo simulations at that time were not able to account for these LEE, because of the limitation on the production and propagation of secondaries below 250 eV for GEANT4 (1 keV for FLUKA).

After the mission, simulations were refined by incorporating the actual solar modulation of the GCR flux observed during the mission elapsed time \citep{bridge,wass23} and extending FLUKA's electromagnetic physics below 1 keV. To achieve this, the Low Energy Ionization (LEI) Monte Carlo program was developed at the University of Urbino Carlo Bo \citep{mattia,bridge,mattia24}. Studies \citep{Grimani21,teorico} demonstrated that ionization is the primary contributor to TM charging, with kinetic emission (low-energy electron emission from surfaces bombarded by keV-MeV particles) and quantum electron diffraction below 100 eV also playing significant roles. Photon-related processes, such as transition radiation, bremsstrahlung, and \v{C}erenkov radiation, were found to be not relevant \citep{Grimani21}. In LEI, low-energy ionization was implemented using the Cucinotta formula \citep{cuci}, while Sakata cross-sections were applied for electrons and positrons \citep{sakata}. Kinetic emission was initially implemented using Schou’s formalism \citep{schou80,Grimani21} but was later found to be overestimated by 30\% \citep{bridge} and thus refined using an \emph{ab initio} approach \citep{taioli23}. The quantum electron diffraction model, initially assuming normal incidence \citep{bridge}, was updated to account for electron direction and Bragg's planes at energies below 100 eV \citep{mattia24}. The TM charging obtained with FLUKA/LEI simulations reproduced nicely the observations carried out with LPF \citep{bridge,mattia24}. 

In 2021 we started under the ESA contract 4000133571/20/NL/CRS ( TEST MASS CHARGING TOOLKIT AND LPF LESSONS LEARNED) the development of  a toolkit (Test Mass Charging Toolkit - TMCTK), GEANT4 based,  for the study of the TM charging for LPF/LISA-like GRS featuring:  the low EM physics learned with FLUKA/LEI (see Section \ref{toolkit}), a general modelization of the particle fluxes, described in Section \ref{spaceenvsec} , and the impact analysis of the charging process on the performance of the mission (see Section \ref{studysection}, \ref{sectiondiscussion} and \ref{sensitivitysection}).
 TMCTK was finally delivered to ESA in 2023 adopting version 11 of GEANT4 released in December 2021 along with an improved version of the GEANT4-DNA module \citep{dna1,dna2,dna3,dna4,dna5}, made to address the simulation of extremely low-energy electromagnetic processes (down to 10 eV) typically used for radiobiology studies \citep{Sakata2018}. 
 
The first aim of this paper is to present the main elements of TMCTK and its first results for LISA mission, that allow to accurately predict the impact on the LISA mission of a wide array of environmental charging scenarios. 
However, despite the improvements, the TMCTK prediction for the charging noise $\lambda_{EFF}$ appear underestimated with respect to FLUKA/LEI and LPF observations. Further investigations on this issue revealed in fact major discrepancies between TMCTK and FLUKA/LEI regarding some particular aspects of LEE propagation, like backscattering and the transmission yields on gold slabs. The details of these discoveries are discussed in Section \ref{sectiondiscussion} of the paper
and will be the subject of future investigation along with the GEANT4 scientific community.

\begin{table}
\renewcommand*{\arraystretch}{1.3}
\caption{Net and effective TM charging and TM equilibrium potential measured with LPF \citep{armano2017b,armanoTM23}.}\label{tab:LPF_Results}
\centering                                      
\begin{tabular}{c|cc}        
\hline\hline                      
& Test mass 1 & Test mass 2\\
\hline  
$\lambda_{NET}$ ( s$^{-1}$) & $+22.9\pm 1.7$ & $+24.5\pm 2.1$\\
$\lambda_{EFF}$ (s$^{-1}$) & $1060\pm 90$ & $1360\pm 130$\\
  $V_{TM}^{EQ}$ (V)& 0.9 $\pm$ 0.3 & 0.96 $\pm$ 0.4\\
\hline
\end{tabular}
\end{table}

\section{LISA test-mass charging Monte Carlo simulation\label{toolkit}}


 \begin{figure}
\centering
\includegraphics[width=0.4\textwidth]{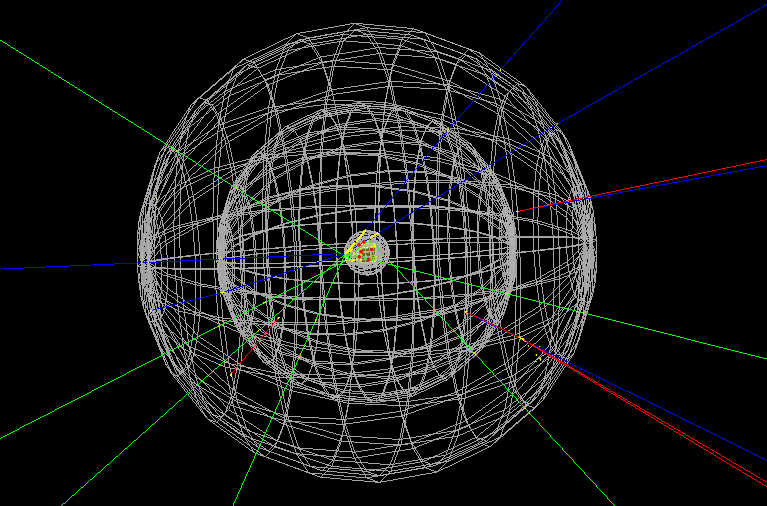}
\includegraphics[width=0.4\textwidth]{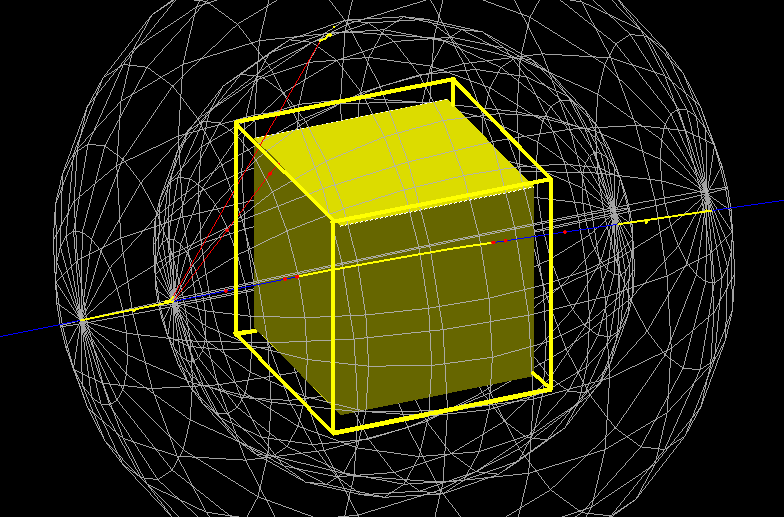}
\caption{Top panel: LISA spacecraft simplified matter distribution around the TMs. Bottom panel: Magnified view of the central part of the geometry modeling the GRS and the TM in GEANT4. The 150 nm wide cubic box modeling the last interface of the EH, as well as the spherical shielding shells, are visualized in wireframe mode for visualization purposes.}\label{fig:geom}
\end{figure}


The estimation of the effects of the space environment on the LISA TM free fall motion is founded on the Monte Carlo simulation of the particle to matter interaction. 
As the LISA spacecraft design is not defined at present,  we have considered a spherical symmetry for the material distribution around the TM consisting of four aluminum concentric spherical shells
sourrounding a gold cubic TM. This material amount is quantified in a total of 16 g/cm$^2$, mainly
concentrated in the vicinity of the TMs. This is an extremely simplified modelization of the matter distribution. The amount of matter was set on the basis of a preliminary design of LISA
spacecraft \citep{geomlisa}. The adoption of a simplified geometry is a common practice\footnote{See for instance \url{https://www.spenvis.oma.be/}} in the design process of space missions to carry out an estimation of the effects of the space environment on sensitive parts of the satellite \citep{vocca2004}.
Moreover, given the importance of emission of secondary particles generated by particle interaction in the immediate vicinity of the TM, the innermost interface
of the GRS EH is modeled as a cubic gold box 150 nm thick and 5.2 cm on the side representing part of the outer film of gold, separated from the TM itself by a 3 mm vacuum gap. This arrangement, the same adopted in \cite{taioli23} allows us to save computation time without losing the significance of the simulation. The thickness of the layer was chosen to be conservatively greater than the mean free path of electrons with energies below 100 eV. Extensive studies of the output of the simulation demontrated the negligible dependence of the results on this parameter.

The particle propagation in GEANT4 occurs in discrete steps properly selected from the active processes. In particular, the active processes characterized by high cross sections for an assigned material and particle energy, have a higher probability to be selected by the simulation engine. The  length of the propagation step depends on the selected physical process, but can also be limited by the user to a custom value.
The particle ionization energy loss is implemented  according to the continuous slowing down approximation (CSDA). After each step, the energy of the simulated particle is lowered accordingly. In general, such approach allows for a reliable description of the production of secondaries and of the particle energy loss over the steplength by maintaining a relatively high efficiency from the point of view of the computation power.

 GEANT4 allows for a modular handling of the different categories of physical processes (electromagnetic, hadronic, decay), that can be independently activated in the simulations.  These modules are included into a "modular physics list" where each process for a given energy and material uses appropriate theoretical models that can be chosen by the users. The activation of a large number of modules enhances the precision of the simulations and the number of possible considered interactions, at the cost of a major increase of the processing time.
The physics list  implemented for this study is the standard QGSP-BIC (\cite{Geant4_PhysicsReferenceManual}), which includes all the principal electromagnetic and hadronic processes, as well as radioactive decay and neutron transport. Of particular interest for us, is the implementation of the module that activates the  electromagnetic interactions. In our  simulation we made  use of the "G4StandardEM-option4" (Opt4) module \citep{opt4}, which is particularly suited for studies where electrons and positrons play a relevant role, allowing us to track these particles down to energies of 100 eV. Opt4 is used in all spacecraft regions except for the outer 150 nm of gold of both TM and EH where the QGSP-BIC physics list was integrated with the GEANT4-DNA module, to account for production and propagation of LEE. The processes implemented in the DNA module will be briefly described in the following subsection.

\subsection{Simulation of low-energy electrons}

GEANT4-DNA replaces the comprehensive multiple scattering processes of Opt4 with more detailed models for LEE interactions, including:
 elastic scattering, electronic excitation,  ionization, vibrational excitation, and molecular attachment.

For electrons in gold, GEANT4-DNA includes specific cross-section models covering energies from 10 eV to 1 GeV. Although GEANT4-DNA can also simulate processes involving nuclei, such as protons and alpha particles, no cross-section models for these particle interactions in gold are currently available. Therefore, custom processes described in Section \ref{PKEsection}  were implemented, analogously to \cite{wass23} to model LEE kinetic emission following the impact of these particles, which, anyways, contribute minimally to the overall LEE production. 

By default GEANT4 allows for the propagation of electrons until their energy is reduced by CSDA below a threshold E$_{min}$, which for GEANT4-DNA module is set by default at 10 eV. 
Below this threshold, the propagation of the particle is stopped.
The discrete interactions have a different low energy threshold E$_{th}$ which encodes the energy validity range of their cross section models. In GEANT4-DNA E$_{th}$ is nominally set at 10 eV as E$_{min}$  .
 However, the minimum energy required for a LEE to escape the gold surfaces of the TM and EH is equal to the gold work function, found to be typically in the range 3.9-5.2 eV \citep{wass23}. As such, in our toolkit, the E$_{min}$ threshold was lowered to 4 eV, without altering (E$_{th}$). This adjustment allows secondary electrons with energies between (E$_{min}$) and(E$_{th}$)reaching the gold-vacuum boundary to escape the surface.

As demonstrated in \citet{taioli23}, this approach effectively reproduces the theoretical and measured electron backscattering yield of gold. Varying the \(E_{min}\) threshold between 3.8 and 4.2 eV showed no significant impact on the $\lambda_{NET}$ and $\lambda_{EFF}$.

\subsection{Low-energy electrons from hadrons}\label{PKEsection}

Kinetic emission of low-energy electrons (LEEs) can also occur when nuclei, such as protons and alpha particles in the keV-MeV energy range, interact with the gold surfaces of test masses (TMs) and electrode housings (EH). Since cross section models for these processes in gold are not included in GEANT4-DNA, two custom processes were implemented in the GEANT4 physics list. These processes simulate the emission of secondary LEEs using a yield-based approach, triggered when protons or alpha particles traverse the gold surfaces.
\begin{figure}[b]
\centering
\includegraphics[width=0.46\textwidth]{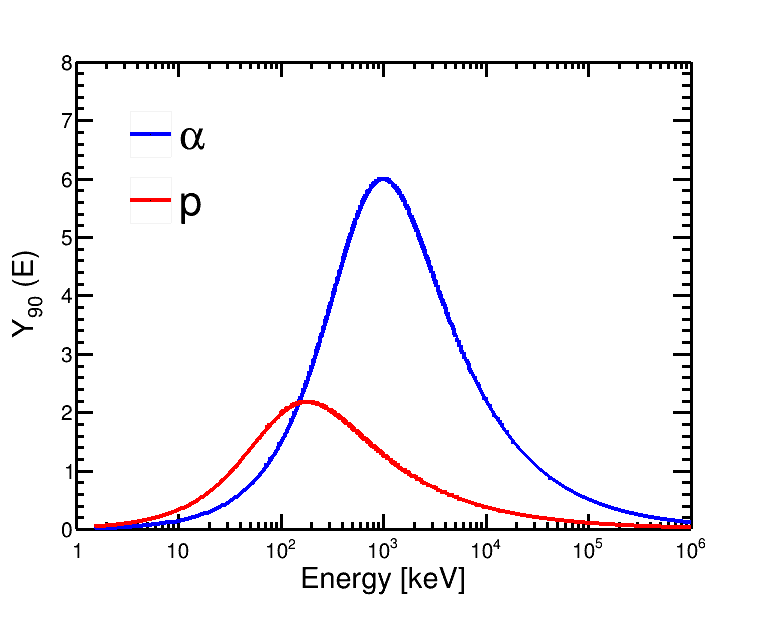}
\caption{ Low-energy electron yield  of the kinetic emission processes for p (red) and $\alpha$ particles (blue) crossing perpendicularly gold surfaces (Y$_{90}$) as a function of p and $\alpha$ energy. 
}\label{fig:PKEteo}
\end{figure}
The number of generated electrons Y(E,$\theta$), as a function of proton energy E and incidence angle $\theta$ with respect to the normal to the surface, is sampled from the average proton and alpha kinetic emission yield function, defined by the following formula \citep{PhysRevSTAB.5.124404,wass23}:
$$
Y(E,\theta) = Y_{max}(\theta)\frac{s E/E_{max}(\theta)}{s -1 + (E/E_{max}(\theta))^2}
$$
where 
$$
E_{max}(\theta) = E_0 [1-0.7(1-\cos(\theta))]
$$
and 
$$
Y_{max} = Y_0[1+0.66 \cos^{0.8}(\theta)].
$$

In the above formulas $Y_0$ is the value of yield for particles normally incident on the surface, $E_{0}$ represents the peak energy of the distribution and s its spectral width.
In the case of protons, values of $Y_0$=2.18, $E_{0}$= 180 keV and s=1.54 have been used \citep{PhysRevSTAB.5.124404}.
For the kinetic emission given by alpha particles, very subdominant with respect to the electron production from  protons, the angular dependence was considered constant, and values of $Y_0$=6, $E_{0}$= 200 keV and s=1.64 were used.
The energy distribution of the LEE produced by these processes is sampled from a narrow ($\sigma$ = 2 eV) distribution centered around 8.4 eV, and their emission direction  is assumed isotropic with respect to the surface normal direction.

Figure \ref{fig:PKEteo} shows the yield functions and the energy spectrum of the LEEs generated by kinetic emission from protons and helium nuclei ($\alpha$ particles) as implemented in the toolkit.

\subsection{Electron Quantum diffraction} \label{QBSsection}

Due to the low energy of  electrons involved in the TM charging process, wave-like quantum mechanical behaviour of these particles, such as diffraction, cannot be disregarded. 

Diffraction occurs when the wavelength of the electron is comparable to the spacing between the atoms in a crystal lattice ($\approx 4$ {\AA} for gold). In order to calculate the probability of diffraction at a given electron energy $E$, one must solve the radial Schr\"odinger equation of an electron into the potential of the crystal lattice:
\begin{eqnarray}\label{eq:rad}
&-\frac{\hbar^2}{2m} \frac{1}{r^2} \frac{d}{dr}\left( r^2 \frac{dR_l(r)}{dr} \right) + \frac{\hbar^2}{2m} \frac{l(l+1)}{r^2} R_l(r) +\\\nonumber
&+\left( -\frac{Ze^2}{r} + V_{ex}(r) \right) R_l(r) = E R_l(r).
\end{eqnarray}
where $Z$ is the atomic number ($Z=79$ for gold), $m$ is the electron mass, $l$ is the angular momentum quantum number and $V_{ex}(r)$ is the exchange potential due to the fact that electrons are identical particles; finally $R_l(r)$ is the radial wavefunction. 

The Schr\"odinger equation \eqref{eq:rad} is solved in two different ranges: inside the potential of a single ion and in the region between atoms, where the potential is constant; the two solutions are matched by imposing that they assume the same value at the boundary of the two regions  and that their derivatives are continuous. With this approach, it is possible to estimate the electron reflection probability. For the details of this procedure see \citet{LEED,Grimani21,mattia24}.


The theoretical calculation described above provides the electron backscattering probability as a function of electron energy and incidence angle with respect to the normal of the gold surface due to quantum diffraction. The GEANT4 toolkit was integrated with this custom process active only in proximity to the two facing surfaces of TM and EH, similarly to the process described in Section \ref{PKEsection}. The incoming electron is reflected back at the same incidence angle according to the estimated quantum backscattering probability, without energy change.

\section{Modeling the space particle environment for LISA}\label{spaceenvsec}

The LISA TM charging will depend on the overall particle flux incident on each spacecraft. 
Given the about 16 g cm$^{-2}$ \citep{geomlisa} of spacecraft material that are expected to shield the TMs on board the three LISA spacecraft, only primary particles above a certain energy threshold will contribute to the charging. This limits are  about 100 MeV/n for hadrons,   
20 MeV for electrons and 100 keV for photons \citep{lisagrb}.
The LISA interferometer is expected to nominally operate  for 4.5 years and up to 10 years, so to fully encompass the variability of environmental conditions during that period
and to evaluate the range of the net  charging and charging noise that plausibly will be observed, we will consider different conditions of the interplanetary medium: long and short-term variations of GCR fluxes and real SEP events of different intensity. 

In general, GCR fluxes and solar particles associated with gradual events \citep{reameslibro,reames22}, lie in the energy range of interest for the TM charging, and their energy dependence must be taken into account to properly assess the TM charging. The composition of  GCR consist approximately of 90\% protons, 8\% helium nuclei, 1\% heavy nuclei and 1\% electrons, where the percentages are meant in particle numbers to the total number \citep{simpson,papini96}. 
Conversely, the particles of SEP events are mainly protons  (99\%), with electrons and nuclei constituting the residual 1\% \citep{reameslibro}.

TMTCK simulates the different fluxes according to a parametrization discussed in the following Sub-sections.

\subsection{Long term variations of Galactic Cosmic Rays}
The cosmic-ray intensity long-term variations ($>$ 1 month) show quasi-eleven and quasi-twenty two year periodicities associated with the solar activity and the global solar magnetic field (GSMF) polarity change. 
During the last three solar cycles 
the overall GCR flux below tens of GeV has been observed to vary  by a factor of four in the inner heliosphere \citep{grim_haspide23,a&aub,a&aub2}. 

LISA is supposed to be launched in 2035 near the maximum of the solar cycle 26 during a positive polarity period of the GSMF, the same of LPF.
In \cite{grim07} it was shown that during positive polarity periods the energy spectra, $J(r,E,t)$, of cosmic
rays at a distance $r$ from the Sun  at a time $t$  are well represented by  the
 symmetric model in the $force$ $field$ $approximation$
 by  Gleeson and Axford \citep[G\&A][]{glax68}. By considering   time-independent interstellar cosmic-ray spectra  $J(\infty,E+\Phi)$ 
and  an energy loss parameter $\Phi$ it is found that:

\begin{equation}
\frac{J(r,E,t)}{E^2-E^2_0}=\frac{J(\infty,E+\Phi)}{(E+\Phi)^2-E^2_0},
\label{eq:ffapprox}
\end{equation}

\noindent where $E$ and $E_0$ represent  the particle total energy and rest mass, respectively.
For Z=1 particles with rigidity (particle momentum  per unit charge) larger than  100 MV, the role of the solar activity is taken into account by 
defining a $solar$ $modulation$ $parameter$ $\phi$ measured in MV that, at these energies, is equal to $\Phi$  \citep{apj1}. 

The proton interstellar spectrum  by \citet{burger2000}  has been adopted  to set the solar modulation parameter estimated by  \citet{uso1,uso2} and used here\footnote{\url{http://cosmicrays.oulu.fi/phi/Phi_mon.txt}}. Unfortunately, in the \citet{burger2000} paper, no helium flux at the interstellar medium is provided. As a result, we have used the \citet{shikaze07} helium interstellar spectrum, selected on the basis  of the BESS experiment data. In order to test the reliability of the model and of 
our approach for  proton and helium flux estimates, we have compared the outcomes of the model to the monthly average space station magnetic spectrometer experiment AMS-02 data gathered in 2016 above 450 MeV/n \citep{ams02_1}. While the model outcome and data for proton flux  show an agreement  within 10\%, the helium flux obtained with the model resulted higher by 25\% with respect to the data. Therefore we normalized the helium flux for LPF in 2016 to the contemporaneous AMS-02 data gathered above 450 MeV/n. For LISA we will consider the same approach since the LISA particle detectors will provide proton and helium differential fluxes up to 400 MeV/n.

\noindent The modulated particle spectra obtained with the G\&A model have been parameterized according to the following equation  \citep{papini96}:
\begin{equation}\label{eq:flux}
F(E) = \frac{A}{(E+b)^\alpha} E^\beta \   \text{Particles}\ (m^2\ sr\ s\ GeV\ n^{-1})^{-1}
\end{equation}

\noindent where the parameter $b$ measured in GeV(/n)  is used to depress the particle flux at low energies and the dimensionless parameters $\alpha$ and $\beta$ allow us  to reproduce  the trend of the spectrum at high energies. Finally, $A$, measured in particles/(m$^2$ sr s (GeV(n$^{-1}$))$^{-\alpha+\beta+1}$), is the normalization constant.
The agreement between equation \ref{eq:flux} with the G\&A model was discussed in detail in \citet{apj2}.

\begin{figure}
\centering
\includegraphics[scale=0.5]{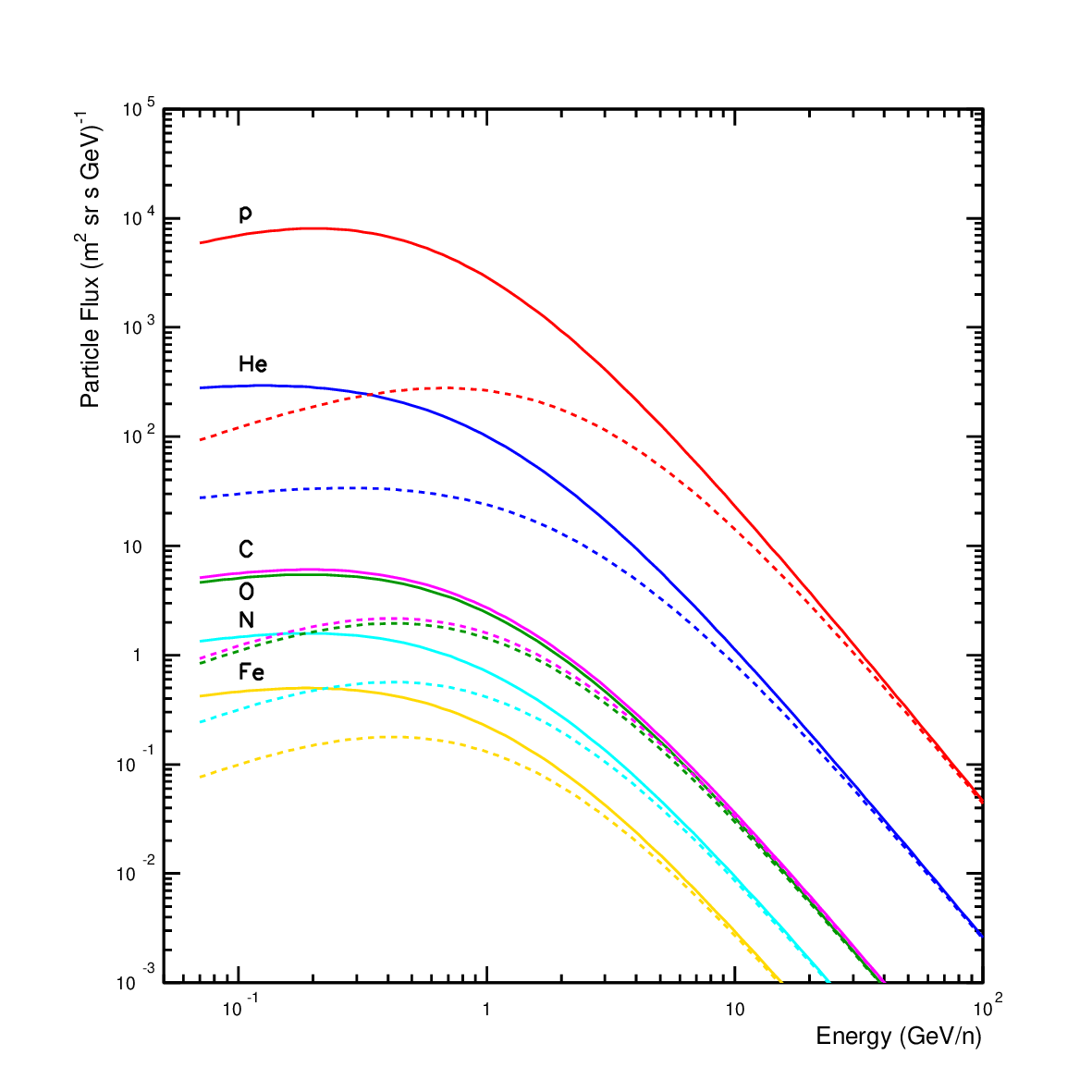}
\caption{From top to Bottom panel: proton (p, red), helium (He, blue), carbon (C, magenta), oxygen (O, green), nitrogen (N, cyan) and iron (Fe, yellow) cosmic-ray energy spectra at solar minimum ($\phi$= 200 MV/c;  continuous lines) band solar maximum ($\phi$= 1200 MV/c;  dashed lines) as extreme conditions for LISA. }\label{fig:gcrspectra_color}
\end{figure}

Since the solar activity at the time LISA will be in orbit is not  known yet, we consider  a solar modulation parameter of $\phi$=200 MV at solar minimum and $\phi$=1200 MV at solar maximum as extreme cases for all particle species on the basis of former observations of the solar modulation parameter. 
In Figure \ref{fig:gcrspectra_color} and Tables \ref{tab:param_p_ecc} our estimates for proton, nuclei  energy spectra  and associated parameterization as from Eqn. \ref{eq:flux} are presented for LISA.

\begin{figure}
\centering
\includegraphics[scale=0.5]{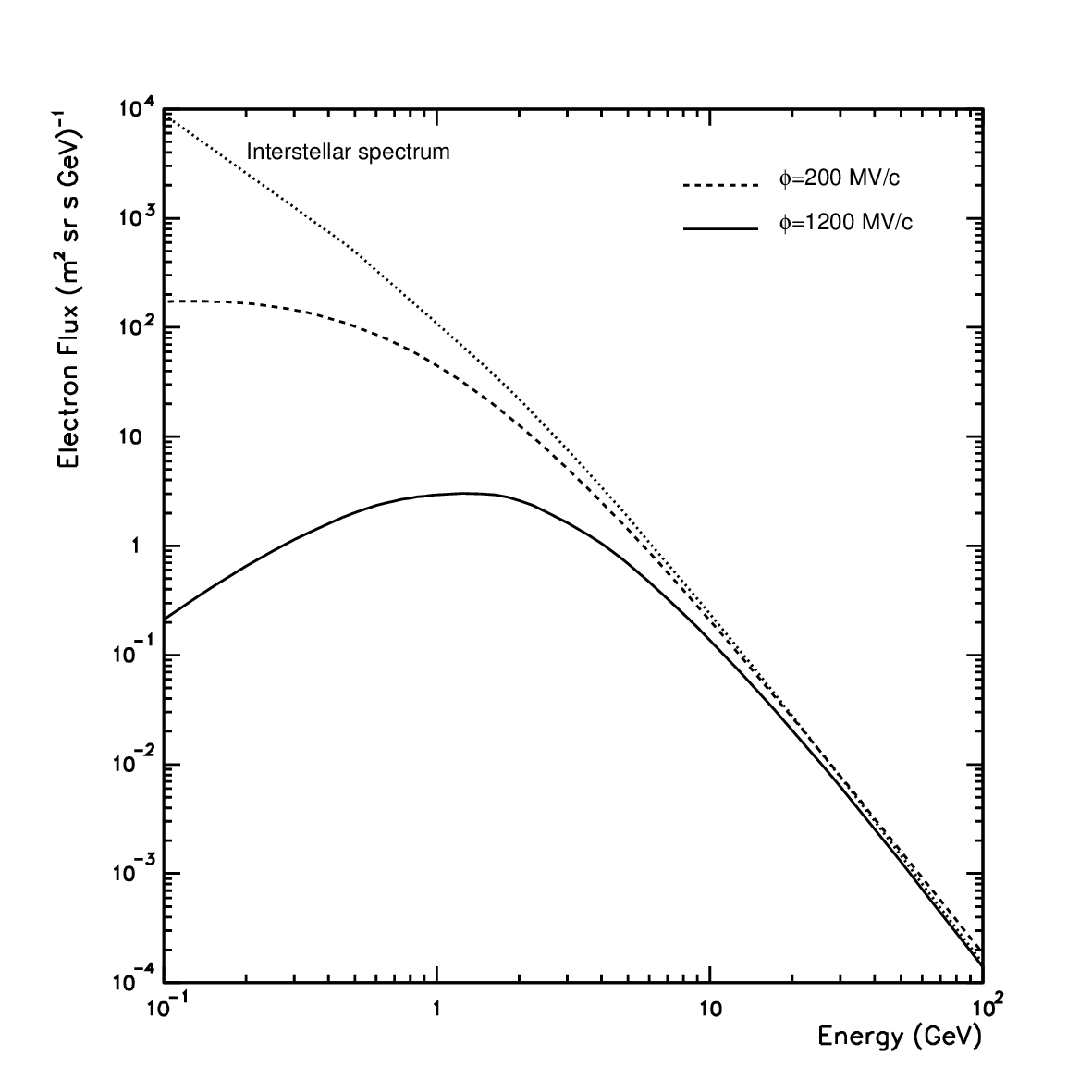}
\caption{\label{fig:gale}Galactic electron energy spectra at solar minimum (dashed line) and solar maximum (continuous line) for LISA. The interstellar spectrum is represented by the top dotted line \citep{moska}.}
\end{figure}

The galactic and interplanetary electron contribution to the LPF TM charging was discussed in  detail \citet{griele}.
Also for LISA we condider only galactic electrons, while electrons below $\simeq$ 20 MeV,   typically of interplanetary and solar origin are disregarded. The electron flux at the interstellar medium by \citet{moska} adopted here (dotted line in Figure \ref{fig:gale}) was found to better reproduce
observations  gathered near Earth at solar minimum, maximum and during different epochs of the GSMF \citep{gri04,gri07}. The G\&A model was also used to modulate the electron interstellar spectrum at 1 AU at solar minimum ($\phi$=200 MV;
dashed line in Figure \ref{fig:gale}) and at solar maximum ($\phi$=1200 MV; continuous line in Figure \ref{fig:gale}).
The parameters in  Eqn. \ref{eq:flux} for electrons are shown in Table \ref{tab:param_e}.

\begin{table*}
\centering
\begin{tabular}{ccccccccc}
& \multicolumn{4}{c}{Solar minimum} & \multicolumn{4}{c}{Solar maximum}\\
\hline
Primary Particle & $A$ & $b$ & $\alpha$ & $\beta$& $A$ & $b$ & $\alpha$ & $\beta$\\
\hline
Protons  & 18000 & 0.65 & 3.66 & 0.87 & 18000 & 2.17 & 3.66 & 0.87\\
Helium   & 850   & 0.99 & 3.10 & 0.35 & 850   & 2.17 & 3.10 & 0.35\\
Carbon   & 28    & 1.05 & 3.25 & 0.50 & 28    & 1.15 & 3.75 & 1.00\\
Nitrogen & 7     & 1.05 & 3.25 & 0.50 & 7     & 1.15 & 3.75 & 1.00\\
Oxygen   & 25.2  & 1.05 & 3.25 & 0.50 & 25.2  & 1.15 & 3.75 & 1.00\\
Iron     & 2.3   & 1.05 & 3.25 & 0.50 & 2.3   & 1.15 & 3.75 & 1.00\\
\end{tabular}
\caption{\label{tab:param_p_ecc}Parameterizations of protons and nucleus energy spectra at 1 AU at solar minimum ($\phi$=200 MV/c) and maximum ($\phi$=1200 MV/c) for LISA. It is worthwhile to point out that A is measured in protons/(m$^2$ sr s GeV$^{-\alpha +\beta +1}$), b in GeV while $\alpha$ and $\beta$ are pure numbers.}
\end{table*}

\begin{table}
\centering
\begin{tabular}{cc}
\multicolumn{2}{c}{Solar minimum}\\
\hline
Energy range & Parametrization \\
\hline
$>20$ MeV & $400 (E+0.82)^{-3.66} E^{0.5}$\\
\hline
\multicolumn{2}{c}{Solar maximum}\\
\hline
Energy range & Parametrization \\
\hline
50 MeV - 1 GeV & $4.5 (E -0.04)^{0.84}$\\
$>1$ GeV & $400 (E+2.5)^{-3.66} E^{0.5}$\\
\hline
\end{tabular}
\caption{\label{tab:param_e}Parametrization of electron energy  spectra above 20 MeV at solar minimum and solar maximum.}
\end{table}


\subsection{Galactic cosmic-ray short-term variations}
Galactic cosmic-ray short-term variations ($<$ 1 month) are observed at the passage of interplanetary structures. In particular,
the GCR flux short-term variations are called recurrent FDs when  associated with the passage of high-speed solar wind 
streams, and non-recurrent FDs (or simply FDs) when are generated by  ICMEs \citep{forbush1,forbush2, forbush3}. 
Detailed studies of the interplanetary physics of GCRs have been carried out with LPF for LISA \citep{apj1,apj2,apj3,mattia23}. 

The recurrent short-term variations show an average duration of 9.1 days  and a marked particle energy dependence below a few GeV, while intense FDs may present an energy dependence up to tens of GeV. The study of the energy dependence of FDs is often difficult because of the contemporaneous presence of solar particles. As case studies, we choose here to consider one recurrent FD observed with LPF and one non-recurrent FD observed in space with the PAMELA experiment \citep{pamFD} and on ground with neutron monitors. The method to calculate the particle fluxparameterization during these events is described in \cite{apj1,apj2,apj3}. It is worthwhile to stress that we were able to use both PAMELA and neutron monitor observations for the proton flux estimate because PAMELA was in orbit near Earth \citep[see][]{villanisp}. 
The evolution of the proton flux during the FD was estimated at the onset (December 14, 2016 15:39 UT), for which we considered the  undisturbed proton flux measured on November 2006 as suggested in \citet{pamFD}, at the  mid-phase at 17:20 UT and at the deep at 24:00 UT of the same day. 

\begin{figure}
\centering
\includegraphics[scale=0.5]{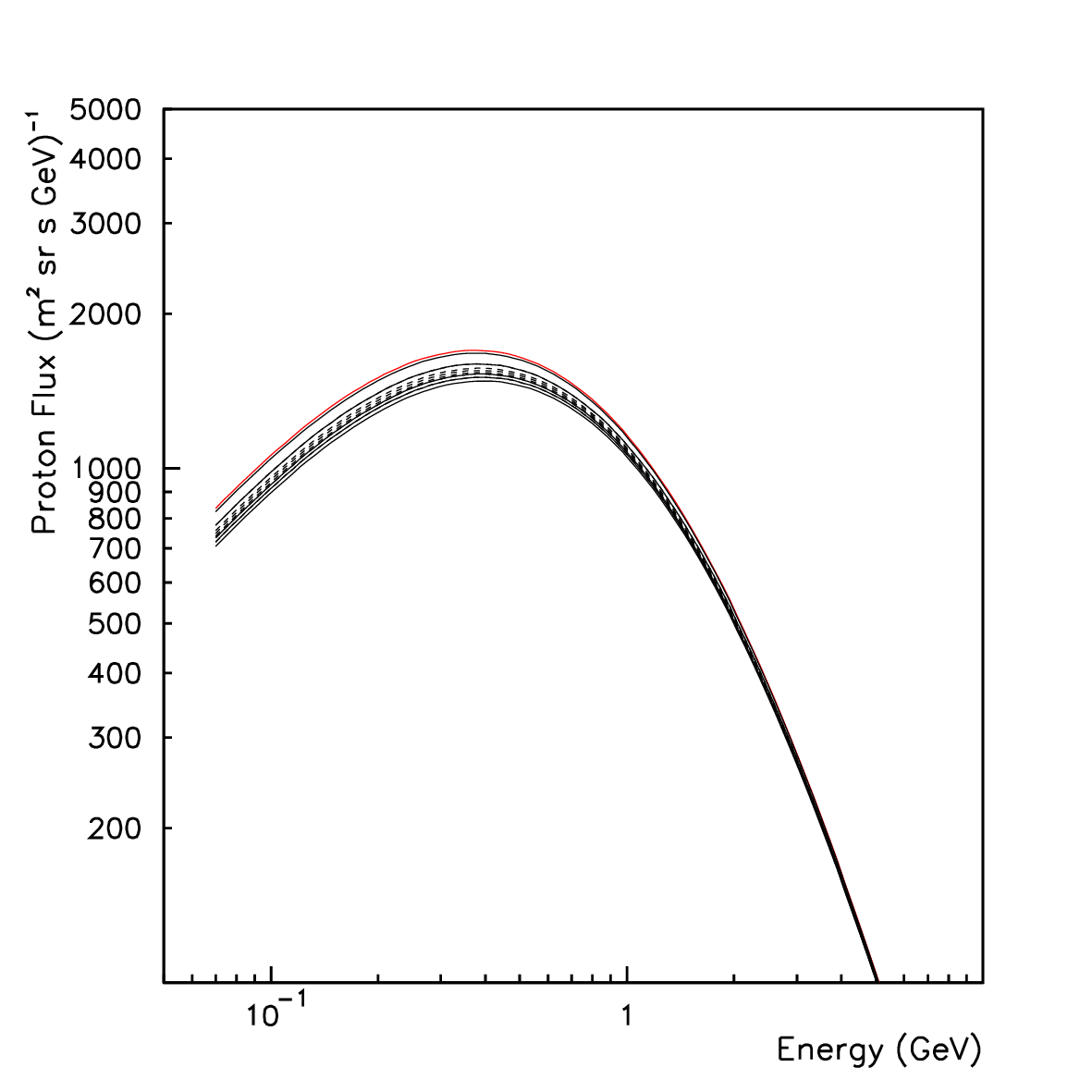}
\caption{\label{RECFD} Recurrent short-term variations of GCRs observed with LPF between November 21, 2016 and December 4, 2016. Continuous lines indicate the decrease phase and the dashed lines the recovery phase. The top red continuous line indicates the proton energy spectrum at the onset of the event.}
\end{figure}

\begin{figure}
\centering
\includegraphics[scale=0.5]{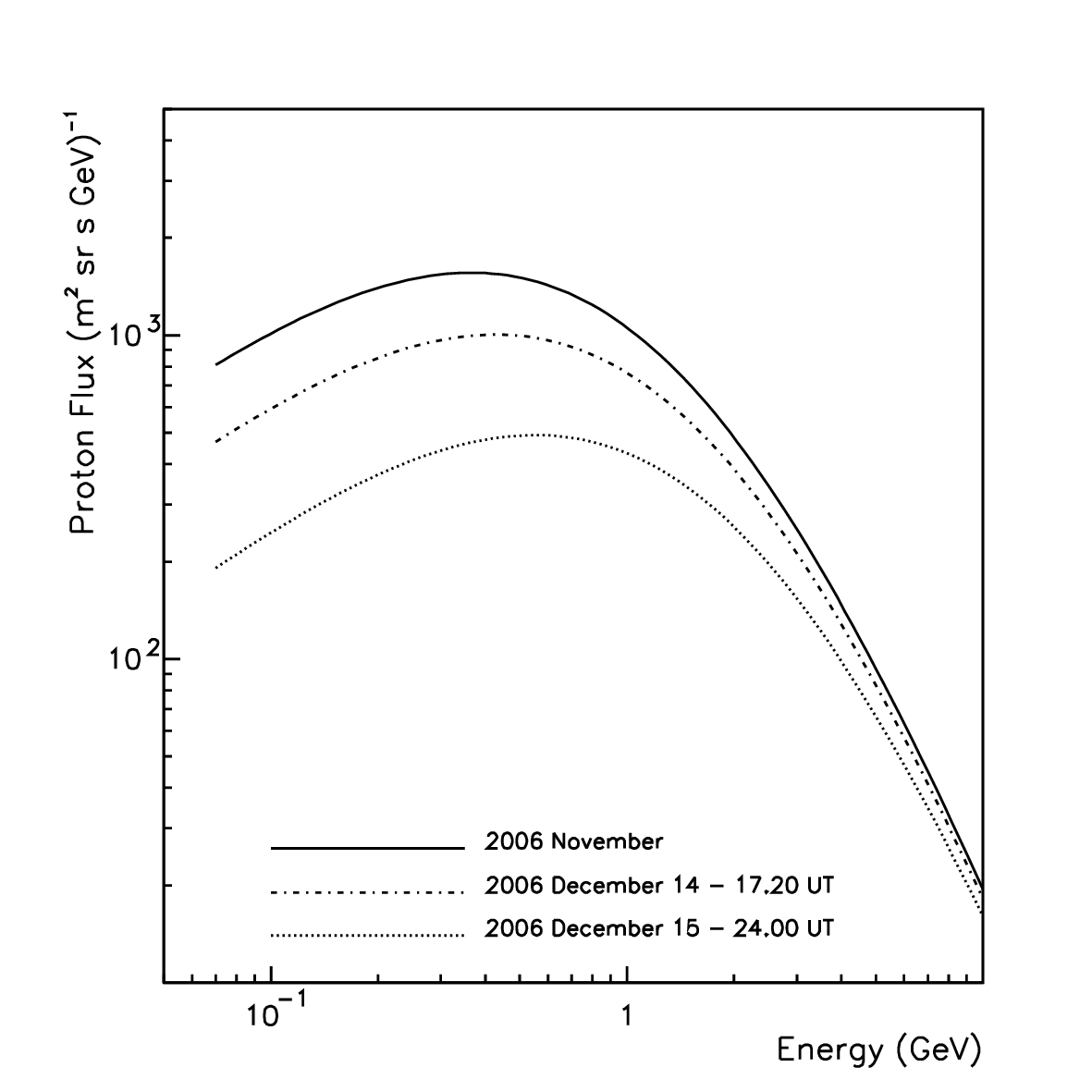}
\caption{\label{NORECFD} Forbush decrease observed on December 14-15, 2006 with the PAMELA experiment in space  and on Earth with neutron monitors.}
\end{figure}

The proton fluxes observed during these events and their parameterization  are reported in Figures \ref{RECFD} and  \ref{NORECFD} and in Table \ref{tab:RECFD} and \ref{tab:NORECFD}.

\begin{figure}
\centering
\includegraphics[scale=0.5]{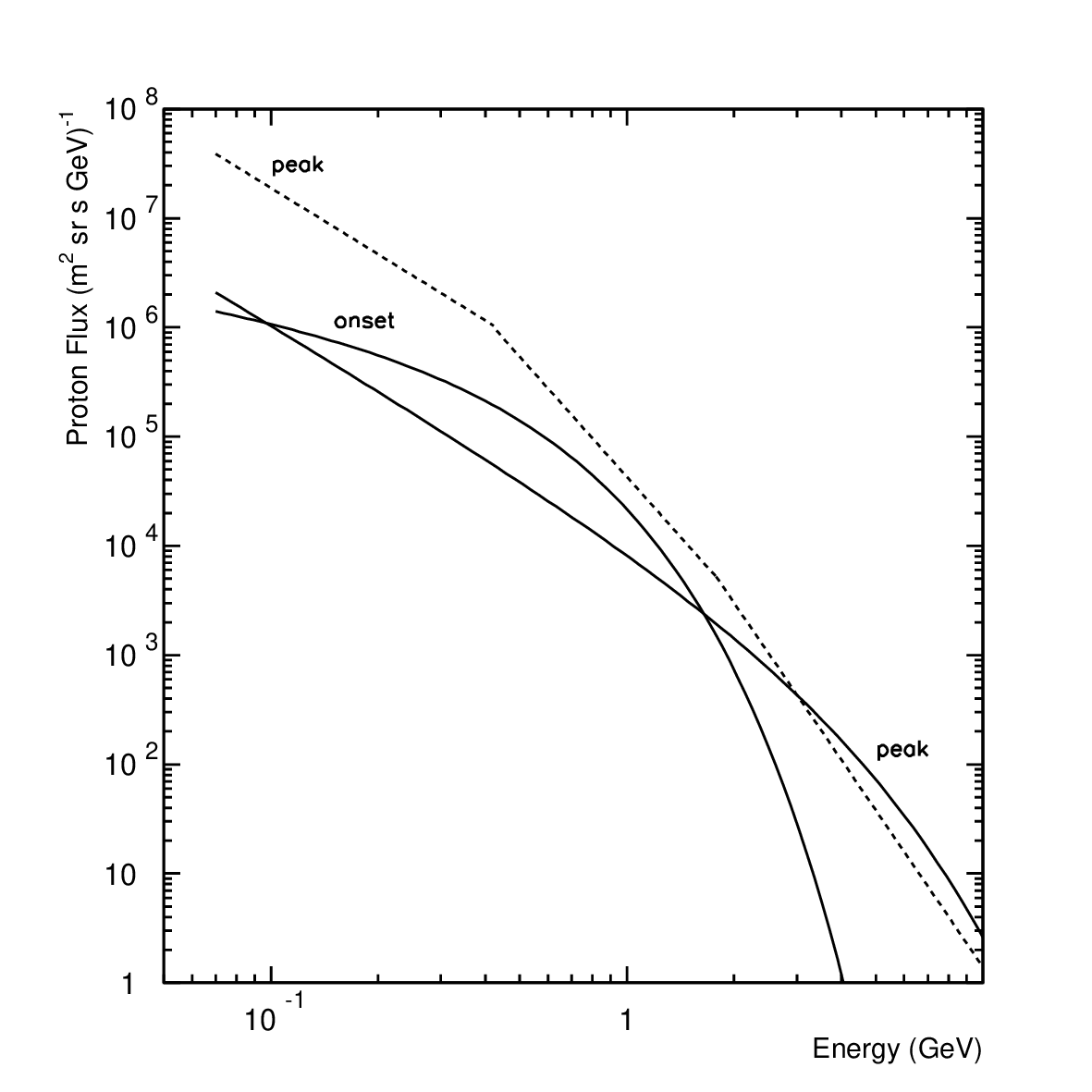}
\caption{Solar energetic proton fluxes observed during the evolution of the gradual events dated September 29, 1989 (dashed line) and 
December 13, 2006  (continuous line). Different phases of the events are indicated. The timings of the events shown in the figure appear in the following in Table \ref{tab:SEP2}. }\label{fig:solp290989}
\end{figure}

\begin{table*}
\centering
\begin{tabular}{ccccc}
\hline
Date & $A$ & $b$ & $\alpha$ & $\beta$\\
\hline
 November 20, 2016 & 18000 & 1.115 & 3.66 & 0.92 \\
 November 21, 2016  & 18000 & 1.12 & 3.66 & 0.92 \\
 November 22, 2016& 18000 & 1.14 & 3.66 & 0.92 \\
 November 23, 2016& 18000 & 1.159 & 3.66 & 0.92 \\
 November 24, 2016& 18000 & 1.165 & 3.66 & 0.92 \\
 November 25-28, 2016 & 18000 & 1.172 & 3.66 & 0.92 \\
 November 29, 2016 & 18000 & 1.165 & 3.66 & 0.92 \\
 November 30, 2016 & 18000 & 1.159 & 3.66 & 0.92 \\
 December 1, 2016& 18000 & 1.156 & 3.66 & 0.87 \\
 December 2, 2016 & 18000 & 1.152 & 3.66 & 0.87 \\
 December 3, 2016 & 18000 & 1.148 & 3.66 & 0.87 \\
 December 4, 2016 & 18000 & 1.14 & 3.66 & 0.87 \\
\end{tabular}
\caption{\label{tab:RECFD}Parameterizations of proton energy spectra during a recurrent variation of the proton-dominated flux measured with LPF between November 20, 2016 and December 4, 2016. It is worthwhile to point out that A is measured in protons/(m$^2$ sr s GeV$^{-\alpha +\beta +1}$), b in GeV while $\alpha$ and $\beta$ are pure numbers.}
\end{table*}

\begin{table*}
\centering
\begin{tabular}{ccccc}
\hline
Date & $A$ & $b$ & $\alpha$ & $\beta$\\
\hline
 November 2006 up to December 14, 2006 15:39 UT & 18000 & 1.17 & 3.66 & 0.87 \\
 December 14, 2006 17:20 UT& 18000 & 1.37 & 3.66 & 0.87 \\
 December 15, 2006 24:00 UT& 18000 & 1.77 & 3.66 & 0.87 \\
 \end{tabular}
\caption{\label{tab:NORECFD}Parameterizations of proton energy spectra during a Forbush decrease observed on December 14-16, 2006. It is worthwhile to point out that A is measured in protons/(m$^2$ sr s GeV$^{-\alpha +\beta +1}$), b in GeV while $\alpha$ and $\beta$ are pure numbers.}
\end{table*}
 \subsection{Solar energetic particle events during LISA}
The charging of the TMs is expected to increase by several orders of magnitude during SEP events \citep{bridge} with respect to the background values associated with the continuous flow of  GCRs.
Being sensitive to hadrons above 100 MeV(/n), it is affected by gradual events characterized by proton acceleration
above 50 MeV. During the evolution of gradual SEP events,  particles show spatial, energy and time variations from onset to decay.
At the onset, the solar particle energy spectra  most likely show a power-law trend with an exponential cut-off, while at the peak
a power-law trend is observed in the majority of cases. 
The proton spatial distribution is characterized by varying pitch angle distributions. It is worthwhile to recall that the pitch angle is
defined as the angle between the particle velocity and the nominal direction of the interplanetary magnetic field Parker spiral. At the onset of the events, the particles 
mainly propagate along the interplanetary magnetic field lines while during  the late phases,  the  arrival direction becomes  isotropic.
Since for the LISA simplified geometry we have considered an  isotropic matter distribution in the  spacecraft, it's not worthy 
to consider the evolution of the  SEP spatial distribution  in the Monte Carlo simulations.
The parameterization of the solar particle energy spectra during the event dynamics was discussed for instance in \citet{grimani2013}.

Solar electrons have been demonstrated to play a minor role in affecting the TM charging and won't be detected by the LISA radiation monitors according to the current design \citep{mazzanti}. As a result, even if solar electrons will reach the LISA S/C before protons due to velocity dispersion,  no short-term forecasting of SEP events will be allowed on board  LISA.


 For events that are magnetically well-connected to the active region of the Sun where the solar eruption occurred, the particle flux on board the three LISA satellites can increase by several orders of magnitude in 15 minutes.
 The average duration of medium-to-strong SEP events (fluence $>10^6$) is about 1.5 days, although some events can last up to 5 days \citep{sha06,HET}.

Unfortunately, from the standpoint of the TM charge measurement, during the LPF operations, no SEP events were observed above the GCR background.
As a result,  for both LPF and LISA the TM charging during SEP events can be only estimated with Monte Carlo simulations \citep{bridge}.
The onset and the peak of gradual SEP events of different intensities are considered here. In particular, 
we study  two SEP events of different fluence: the peak of the event dated September 29, 1989 \citep[fluence 10$^7$-10$^8$ protons cm$^{-2}$,][]{miroshi2000} and the event observed by the PAMELA experiment in space on December 13, 2006 \citep[fluence between 10$^6$ and 10$^7$ protons cm$^{-2}$][]{pamFD}. 
The  proton fluxes observed during these events are shown in Figure \ref{fig:solp290989}. 
Data from the Solar Orbiter mission 
\citep{Daniel,Cesar,HET}\footnote{https://soar.esac.esa.int/soar/} on SEP event occurrence above 70 MeV will allow us in the near future to investigate in detail the effects of several other event dynamics on LISA.

We stress that SEP events with  fluences of 10$^5$-10$^6$ protons cm$^{-2}$ are not observed at solar minimum above the background of GCRs above 70 MeV. 

On the basis of observations gathered during the past solar cycles,
the expected number of SEP events  during the LISA operations can be  at most  of the order of 10 per year during the first part of the mission since the LISA launch is scheduled  at the maximum  of the solar cycle 26 \citep{singh19}.

\section{LISA Test Mass charging case studies}\label{studysection}


\begin{table*}[ht]
\centering
\begin{tabular}{cccccccccc}
&& \multicolumn{4}{c}{Solar minimum} & \multicolumn{4}{c}{Solar maximum}\\
\hline
Primary Particle & Sim. time  & $\lambda_{NET}$ & error & $\lambda_{EFF}$ & error & $\lambda_{NET}$ & error & $\lambda_{EFF}$ & error\\

 & [s] & [s$^{-1}$] & [s$^{-1}$] & [s$^{-1}$] & [s$^{-1}$] & [s$^{-1}$] & [s$^{-1}$] & [s$^{-1}$] & [s$^{-1}$]\\
\hline
& && & & & & & &\\
Protons  &10000  &51.50 & 0.80 & 440.5 & 8.79  & 4.30 &0.14  & 154.70 &15.50 \\
Helium   &50000  &5.60    & 0.20 & 77.00 & 2.10 & 1.05 & 0.05 & 53.30 & 7.39\\
Carbon   &50000  &0.50    & 0.10 & 17.50& 1.10&0.12 &0.01 & 6.50 & 0.10 \\
Nitrogen &50000  & 0.25    &0.05 & 2.0 & 0.1 & 0.12 & 0.01& 0.5 & 0.01 \\
Oxygen   &50000  &0.30  & 0.05& 14.50& 2.00&0.13&0.01&8.18&0.49  \\
Iron     &150000  &0.20  & 0.02&  30.50&  2.50&0.05 &0.01 &15.60 &2.30\\
electrons&100000  &-0.50   & 0.05&  69.05&  1.70& 1.00& 0.02 &28.90 & 1.85\\
& && & & & & & &\\
Total & - & 57.85 & 0.83 & 651.05 & 9.80 & 6.77 & 0.15 & 267.70& 17.40\\
\\
\end{tabular}
\caption{Monte Carlo estimates of $\lambda_{NET}$ and $\lambda_{EFF}$ for the different  species of the GCR particles during $\phi = 200$ MV (Solar Minimum) and $\phi=1200$ MV (Solar Maximum) conditions.\label{tab:longterm}}
\end{table*}

\begin{table*}[ht]
\centering
\begin{tabular}{ccccccc}
Date & Event & Phase & $\lambda_{NET}$ & error & $\lambda_{EFF}$ & error\\
 &  &  & [s$^{-1}$] & [s$^{-1}$] & [s$^{-1}$]& [s$^{-1}$]\\
\hline
\\
14:12 UT Sept 29, 1989 & SEP & Peak &13414  & 75 & 40912 & 1781  \\
03:18 UT Dec 13, 2006 & SEP &Onset &2351  & 50 & 6038 & 270  \\
04:33 UT Dec 13, 2006 & SEP & Peak   &1037  &35    & 2519 & 110   \\

 15:39 UT Dec 14, 2006 & FD & Onset & 13.6  & 0.5   & 262  & 55   \\
 17:20 UT Dec 14, 2006 & FD & Mid-Phase   & 9.8  & 0.3    &  187 & 16    \\
00:00 UT Dec 15, 2006 & FD & Dip-Phase &  6.1 & 0.2  &  163 & 17   \\
\\
\end{tabular}
\caption{\label{tab:SEP2} Monte Carlo estimates of $\lambda_{NET}$ and $\lambda_{EFF}$ for proton fluxes under non-recurrent transient solar events.}
\end{table*}

\begin{table}[ht]
\centering
\begin{tabular}{ccccc}
& \multicolumn{2}{c}{Solar minimum} & \multicolumn{2}{c}{Solar maximum}\\
\hline
Particle & $\lambda_{NET}$ & $\lambda_{EFF}$ & $\lambda_{NET}$ & $\lambda_{EFF}$ \\
  & [s$^{-1}$]& [s$^{-1}$] & [s$^{-1}$] & [s$^{-1}$] \\
\hline
\\
Protons  & 48.7 & 672.1 & 14.1 & 255.9 \\
Helium   & 15.0 & 289.8 & 4.1  & 183.1 \\
Carbon   & 1.4  & 93.9  & 0.8  & 57.5  \\
Nitrogen & 0.4  & 26.9  & 0.2  & 21.1  \\
Oxygen   & 1.7  & 123.9 & 0.9  & 64.6  \\
Iron     & 0.4  & 72.8  & 0.3  & 63.3  \\
electrons& -3.2 & 216.7 & -0.1 & 91.1  \\
\\
Total    &  64.4& 1496.1& 20.3 & 737.6 \\
\\
\end{tabular}
\caption{FLUKA/LEI Monte Carlo estimates of $\lambda_{NET}$ and $\lambda_{EFF}$ for  different  species of  GCR particles  during $\phi = 200\,$ MV and $\phi=1200\,$ MV solar conditions \citep{mattia24}.}\label{tab:resultLEI}
\end{table}

\begin{table}[ht]
\centering
\begin{tabular}{cccc}
Date & Phase & $\lambda_{NET}$  & $\lambda_{EFF}$ \\
  &  & [s$^{-1}$]  & [s$^{-1}$] \\
\hline
\\
December 13, 2006  & ONSET  &2425  & 5695 \\
December 13, 2006  & PEAK   &1123  & 2360 \\
September 29, 1989 & PEAK   &16358 & 56733 \\  
\\
\end{tabular}
\caption{FLUKA/LEI Monte Carlo estimates of $\lambda_{NET}$ and $\lambda_{EFF}$ for proton fluxes during SEP events.}\label{tab:SEPLEI}
\end{table}


In space, the deposit of a charge on the TM is an inherently Poissonian process. Each deposit $j$, say one elementary charge or more, of either signs, has its rate $\lambda_j$.
The TM charges up with a rate: 

\begin{equation}\label{lnet}
    \lambda_{NET} = \sum_{j=-\infty}^{+\infty} j\,\lambda_j \quad (\text{s}^{-1}), 
\end{equation}

This charging process produces on the TM shot noise $S_{\dot{Q}}=2eI$ with the current $I = \lambda_{EFF}e$, where:

\begin{equation}\label{leff}
    \lambda_{EFF} = \sum_{j=-\infty}^{+\infty} j^2\,\lambda_j \quad  (\text{s}^{-1}).
\end{equation}

The integral of the charging yields the "red" power spectral density of the deposited 
 charge $S_Q=2\lambda_{EFF}e^2/(2 \pi f)^2$, as a function of the frequency $f$ \citep{araujo}.




The LISA observatory measures the relative motion of TM pairs in the 0.1 mHz to 1 Hz band \citep{amaro2017,lisa_redbook}. Therefore,  the calculation of the $\lambda_{NET}$ and $\lambda_{EFF}$ must be representative at time scales of 1000s seconds or more for the galactic cosmic-ray flux.

\begin{figure}
\centering
\includegraphics[width=0.5\textwidth]{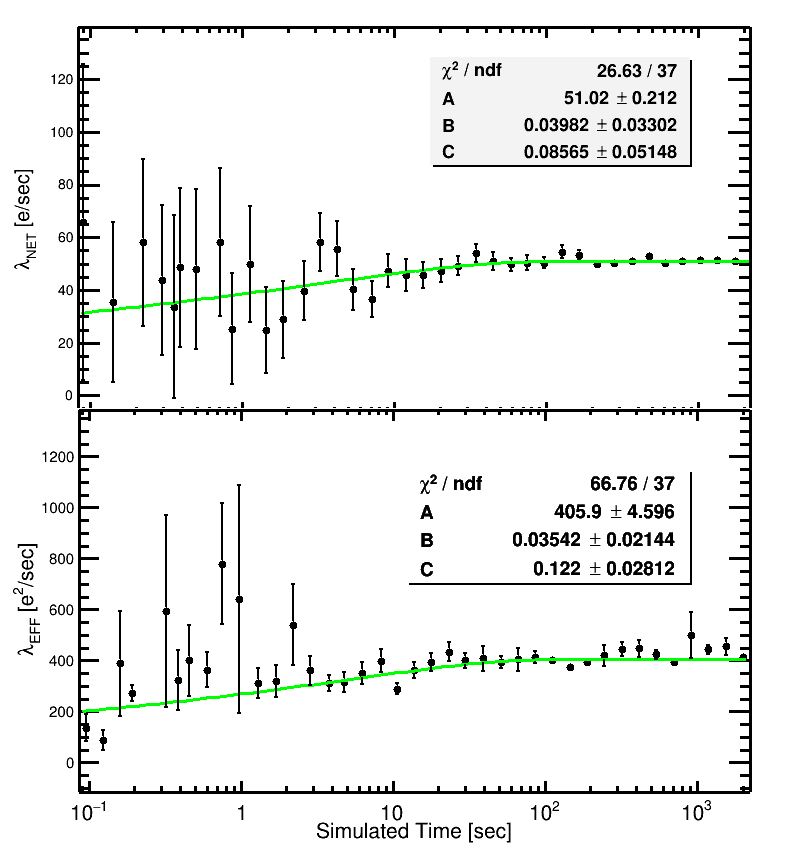}
\caption{Evolution of $\lambda_{NET}$ and $\lambda_{EFF}$ over simulation time for the proton energy spectrum at solar minimum ($\phi$ = 200 MV).  Error bars indicate statistical uncertainty. 
The green line represents an exponential fit to data to show the trend (A(1-e$^{-Bt}$)$^C$). }\label{fig:lambdatimeevo}
\end{figure}

Figure \ref{fig:lambdatimeevo}  illustrates the evolution of  $\lambda_{NET}$ and $\lambda_{EFF}$ as a function of the simulation time, for the proton flux at solar minimum ($\phi$ = 200 MV). 
The error bars on the figure represent the statistical uncertainty on the results. As expected, the uncertainties on $\lambda_{NET}$ decrease uniformly with simulation time, while showing an irregular trend on $\lambda_{EFF}$. This is due to the fact that, given the quadratic dependence of the latter on the deposited charge, the occurrence of events with high charging (large deposits) has a more marked relevance on the evaluation of the uncertainty. 
Both $\lambda_{NET}$ and $\lambda_{EFF}$ show an increasing trend, saturating to a constant value above $\sim$ 100 second simulation time. This is expected: because of the power-law cosmic-ray spectral shape, not enough high-energy (>10 GeV/n) particles are simulated for low simulation times.
These high-energy particles are more likely to be associated with a large production of secondaries deposited on the TMs.
We fitted the $\lambda_{NET}$ and $\lambda_{EFF}$ trends with an asymptotic exponential function, obtaining the results shown in the same Figure, where the asymptote (A) represents the expected value of the two parameters.
From the figure it is clear that a good convergence of the results, expecially for $\lambda_{EFF}$, can be obtained only with simulation times above 10$^3$ sec for protons. We set conservatively to 10$^4$ s the simulation time for protons at both solar minimum and maximum. This minimum simulation time was further increased for rare particles in cosmic rays such as nuclei and electrons. The simulation time used for the various species are reported in the second column of Table \ref{tab:longterm}.

\begin{figure}
\centering
\includegraphics[width=0.5\textwidth]{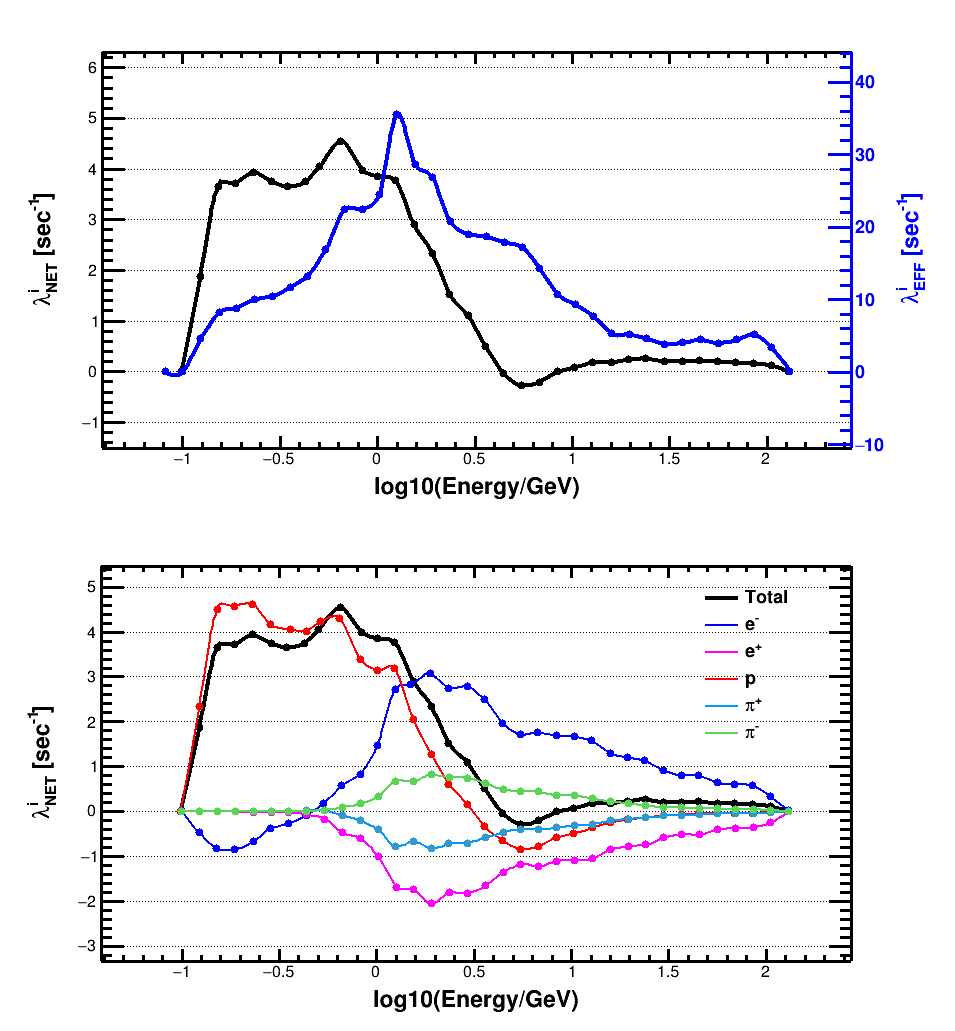}
 \caption{Top panel: Comparison of the contribution to total $\lambda_{NET}$ and to total $\lambda_{EFF}$ for a primary proton flux ($\phi$ = 200 MV) and particles of different energies. Bottom panel:  Break down of the $\lambda^{i}_{NET}$ from the most abundant particle species contributing to the total $\lambda_{NET}$ as a function of the energy of the primary proton. }\label{fig:chargingcontrib}
\end{figure}

In the top panel  of Figure \ref{fig:chargingcontrib}, as an example, we have reported the contributions to $\lambda_{NET}$ and $\lambda_{EFF}$ as a function of the energy of primary protons at solar minimum. The  $\lambda_{NET}$ decreases abruptly above 1 GeV. The meaning of this sharp decrease  can be inferred by observing the break-down of the contribution of the different secondary particle species generated by the interaction of the primary protons appearing in the bottom panel of the same figure. Below 200 MeV the $\lambda_{NET}$ is dominated by protons, stopping in the TM. The production of positive and negative charged  pions (shown in the figure) above this energy tags the occurrence of hadronic interactions in the body of the TM (minimum traversed gold material grammage 91 g cm$^{-2}$).
The secondary particles escape the TM in a larger number with respect to those penetrating from outside: escaping negative charged particles charge positively the TM, the opposite occurs for positively charged particles.
However, the electron curve is also affected by an extra production of electrons by particle ionization. It is worthwhile to point out that globally positive and negative pions, electrons and positrons give a negligible contribution to the total charging with respect to protons.
\begin{figure*}
\includegraphics[width=\textwidth]{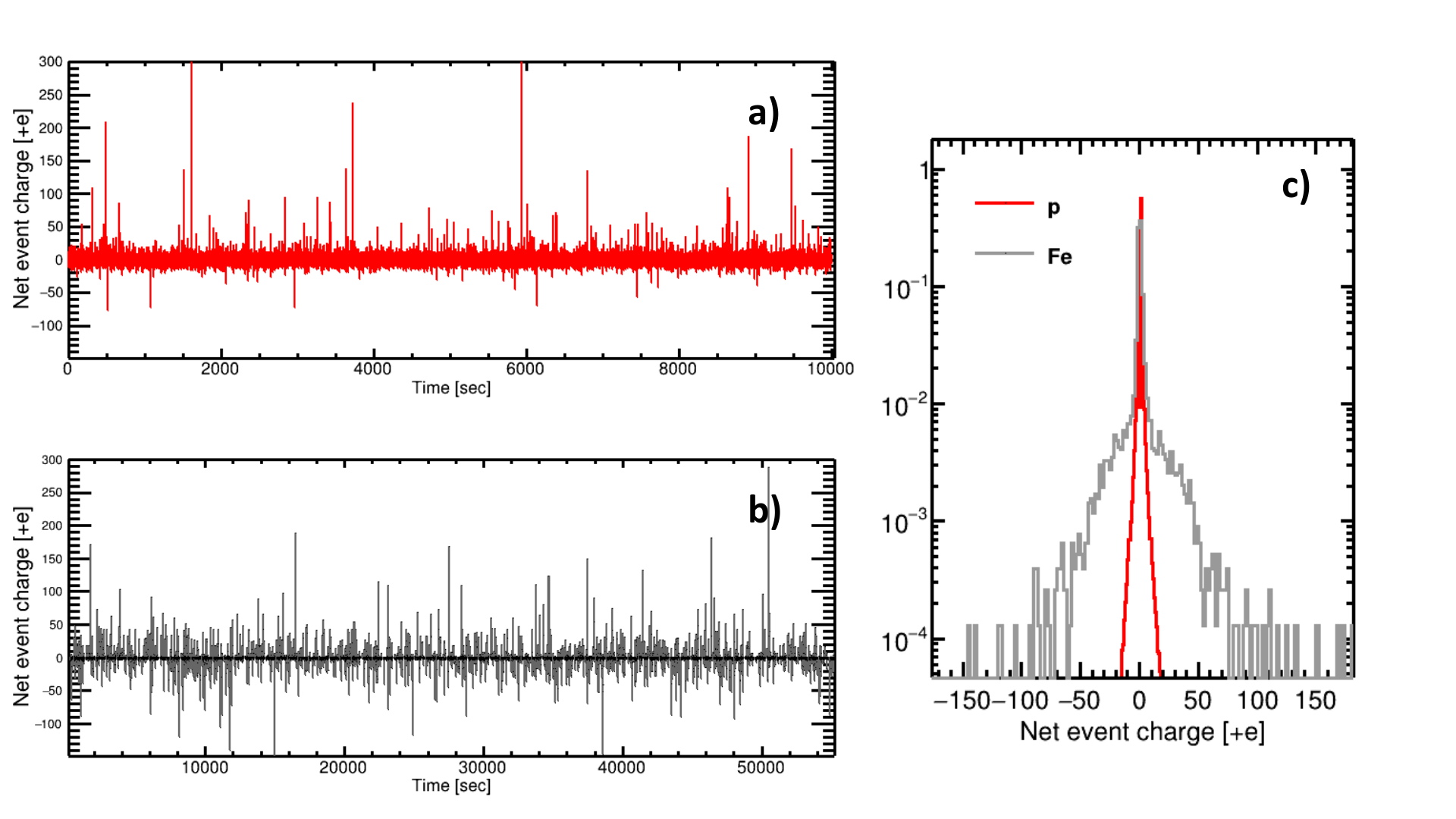}
\caption{Comparison of charging time series for proton  (a) and iron nuclei (b) fluxes at $\Phi = 200$ MV. Panel (c) shows a comparison between the charging histograms of the two time series normalized to the number of events.}\label{fig:timeseries}
\end{figure*}

The trend of  $\lambda_{EFF}$ as a function of the primary proton energy shows contributions peaking for primary proton energies in the GeV range and with significant contributions spanning from 150 MeV to 50 GeV (see Figure \ref{fig:chargingcontrib}). Charge deposits of opposite signs sum up quadratically in the  $\lambda_{EFF}$ formula (Eqn. \ref{leff}) and do not cancel out.

The discussions reported above, allowed us to set properly the simulation timeframe for all particle species sensibly contributing to the TM charging as discussed in Section \ref{spaceenvsec}. 
As an example of TM charging timeseries, we report in Figure \ref{fig:timeseries} the Monte Carlo simulation results obtained for primary protons and iron nuclei under the same solar modulation conditions ($\phi$=200 MV).   
 Both timeseries exhibit a prevalence of positive charge deposits reaching values of several hundreds of elementary charges per second deposited on the TM. As expected, these events are significantly more frequent in the iron case, due to the overall amount of electrons generated by ionization which scales with the square of the particle charge.

\begin{figure}[h]
\centering
\includegraphics[width=0.46\textwidth]{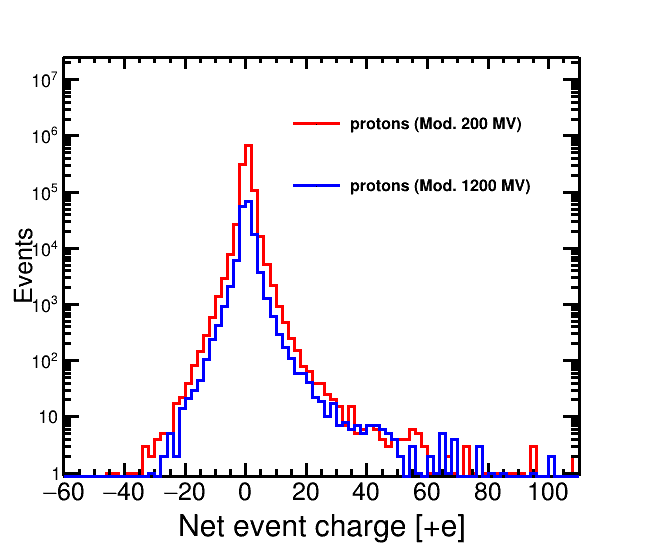}
\includegraphics[width=0.45\textwidth]{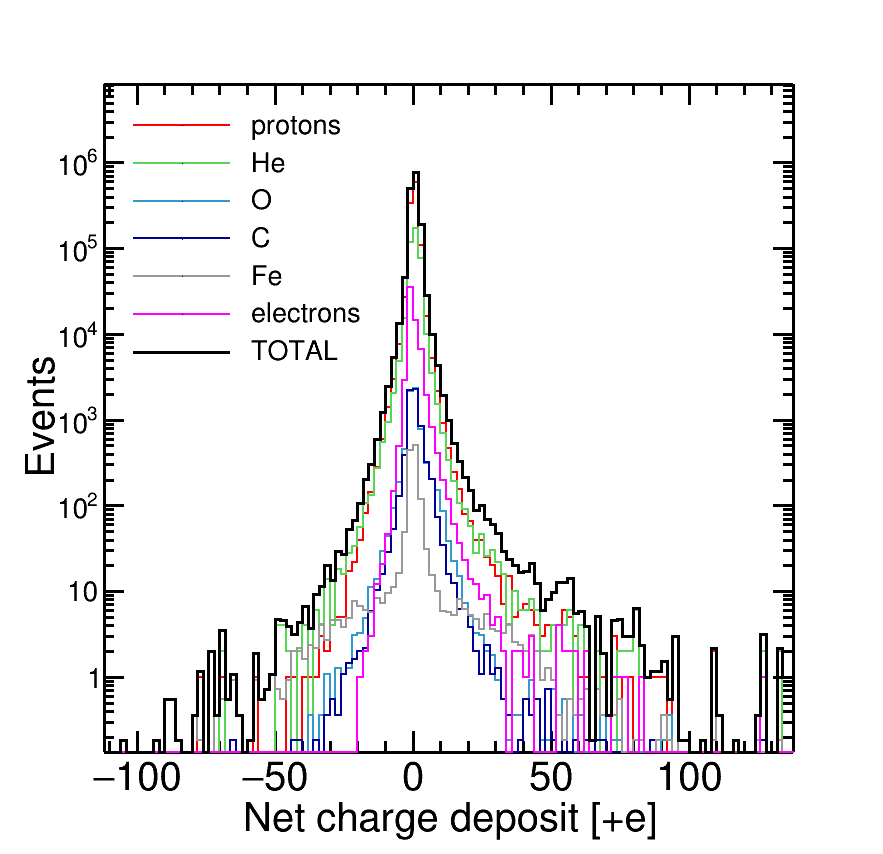}

\caption{Top panel: Comparison of charging histograms obtained from simulations of proton fluxes at extreme solar minimum ($\Phi = 200$ MV) and solar maximum ($\Phi = 1200$ MV) conditions.
Bottom panel: Total charging histogram obtained from complete simulation of CR flux (10000 seconds) at solar minimum condition ($\Phi$=200 MV), broken down in its contributions from every simulated CR specie.}\label{fig:charginghistograms}
\end{figure}

\begin{figure}
\centering
\includegraphics[width=0.5\textwidth,height=6.3cm ]{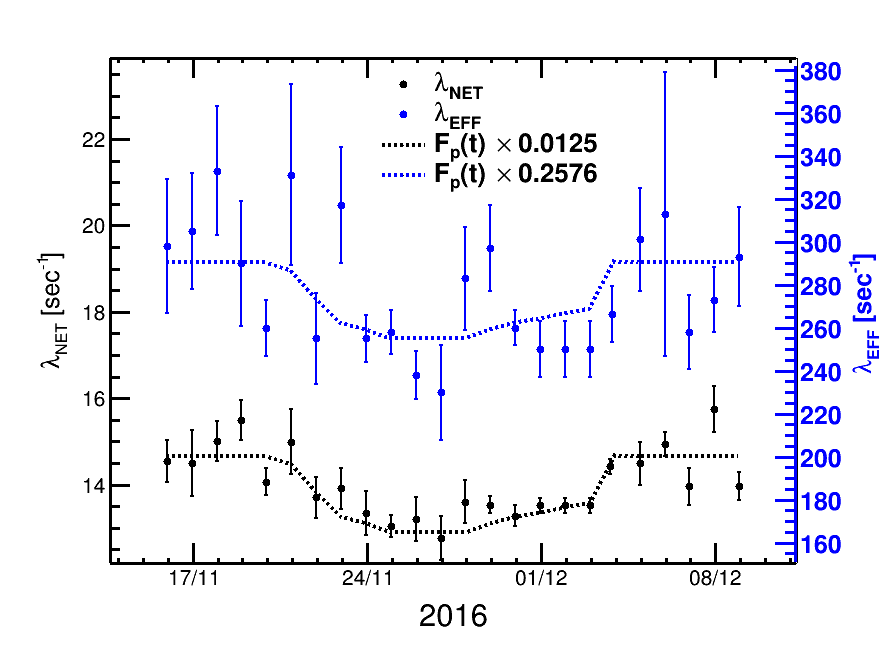 }
\includegraphics[width=0.52\textwidth,height=6.3cm]{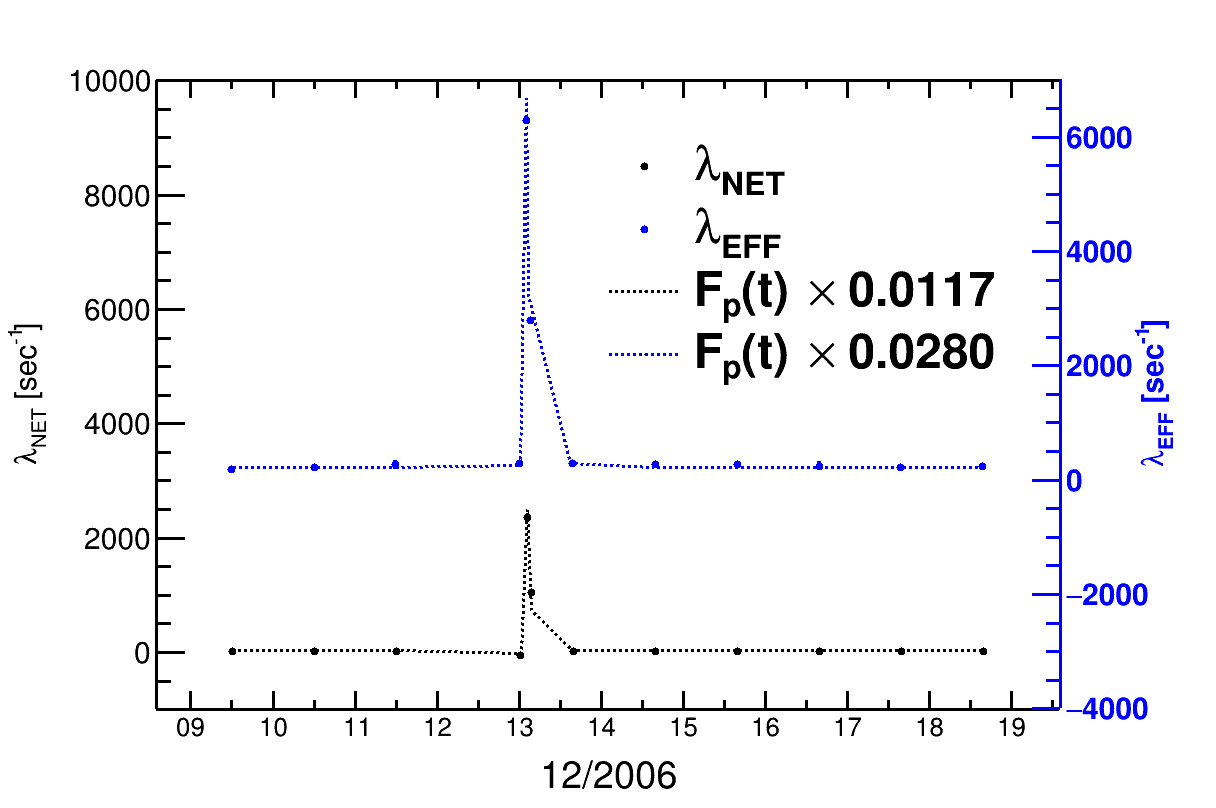 }
\includegraphics[width=0.51\textwidth,height=6.3cm]{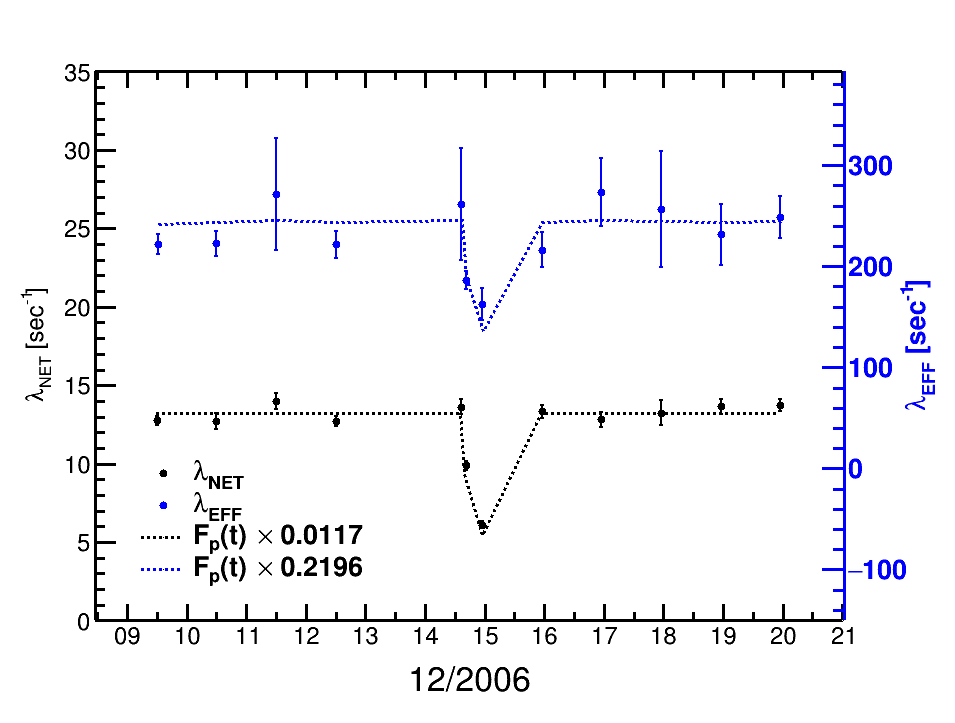 }
\caption{Top panel: Net (black points) and effective (blue points) charging of the LISA TM during the recurrent GCR variation associated with the passage of one high-speed solar wind stream observed with LPF in November 2016. Only the proton flux was simulated. As a comparison, the trend of the simulated proton flux $F_{p}(t)$, integrated in the 0.5-1.5 GV rigidity bin, was superimposed to the $\lambda_{NET}$ and $\lambda_{EFF}$ data through multiplication by a constant (dashed lines).
Central panel: Net (black points) and effective (blue points) charging of the LISA TM during the SEP event of December 13, 2006.The dashed lines show the same comparison of the top panel with the normalized proton flux.
Bottom panel:  Net (black points) and effective (blue points) charging of the LISA TM during the Forbush decrease following to the SEP event of December 13, 2006 (SEPs were subtracted from data).The dashed lines show the same comparison of the top panel with the normalized proton flux.}
\label{fig:monthevo}
\end{figure}
 This evidence emerges more clearly in the  
 \textit{charging histograms} \footnote{Events with 0 deposit are not included in the charging histograms}, shown in the right panel of the same figure, where one can notice that the occurrence of large charge deposits is consistently higher for the iron nuclei.
 
 Figure \ref{fig:charginghistograms} (top panel) presents a comparison between the charging histograms obtained from two proton simulations at extreme solar minimum ($\phi = 200$ MV) and solar maximum ($\phi = 1200$ MV) conditions.
 The deposit rates (see top panel of Figure \ref{fig:charginghistograms}) are consistently lower for solar maximum fluxes. This inbalance fades out for higher and higher charge deposits since solar maximum fluxes are relatively poor in low-energy particles with respect to solar minimum fluxes.

 A global charging histogram from a complete GCR flux simulation of 10000 seconds at extreme solar minimum ($\phi$ = 200 MV) is presented in Figure \ref{fig:charginghistograms} (bottom panel), broken down by contributions from each simulated particle species. Only species with the highest expected contributions were simulated, using the flux parametrization discussed in Section \ref{spaceenvsec}. As can be seen, the peak of the distribution, which is mostly determining the net charging, is dominated by proton and helium contributions, while the tails, mostly affecting $\lambda_{EFF}$ are populated by other particle species in a more than proportional way with respect to their relative abundance.

 
 The results regarding the $\lambda_{NET}$, $\lambda_{EFF}$ and their respective uncertainties obtained with TMCTK in different space environment conditions and for different particles are reported in Tables \ref{tab:longterm}.
 From this table it appears clear that the dominant contribution to the total $\lambda_{NET}$ and $\lambda_{EFF}$ is due to primary protons. The contribution of nuclei is in general smaller, but relatively important with respect to the abundance of these particles in GCRs. This is particularly evident for iron nuclei, which give a contribution of $\sim$ 10\% to the total $\lambda_{EFF}$ due to the combined effects of its high Z and relative abundance with respect to other heavy nuclei. 
 Also electrons, constituting only 2\% of the CGR sample, give a $\sim$ 10\% contribution to the total effective charging. Moreover, it is the only particle species showing a marginally negative contribution to $\lambda_{NET}$. 
 
 In Figure \ref{fig:monthevo} (top and bottom panel) and Table \ref{tab:SEP2}, we present the time evolution of $\lambda_{NET}$ and $\lambda_{EFF}$ for the recurrent short term variation appearing in Figure \ref{RECFD} and for the non-recurrent shown in Figure \ref{NORECFD}.  
 In Table \ref{tab:SEP2} we have reported also the results on the simulation of two SEP events dated September 29, 1989  and December 13, 2006 (see Figure \ref{fig:solp290989}).
 For SEP events, we carried out simulations for a fixed duration of 6 seconds given their rapid evolution and their extremely high flux with respect to GCRs.  Figure \ref{fig:monthevo} (middle panel) shows also the time evolution of $\lambda_{NET}$ and $\lambda_{EFF}$ for the December 13, 2006 SEP.
The $\lambda_{NET}$ appears to follow the variations of the input fluxes being dominated by the primaries stopping inside the TM. This correlation is less evident for $\lambda_{EFF}$, which is more influenced by events producing a large number of secondary particles.

\section{Discussion of results}\label{sectiondiscussion}

The results of the charge rate $\lambda_{NET}$ from protons in Table \ref{tab:longterm} appear in relatively good agreement with those presented in \cite{armanoTM23} obtained with a Monte Carlo simulation using GEANT4 version 10.3 (without the DNA module), where the ionization is treated using Opt4 physics,  and the LEE production from electrons and hadrons is managed with a yield-based approach. In \cite{armanoTM23} the net charging for the LPF TM was calculated on the basis of the INTEGRAL mission measurements by considering the minimum of the  solar cycles 23 and the maximum of the solar cycle 24: the estimates for $\lambda_{NET}$ range from +$8$ s$^{-1}$ to +$40$  s$^{-1}$ , whereas we find +$4$ s$^{-1}$ to +$51$ s$^{-1}$ under extreme conditions of solar maximum and solar minimum (see Table \ref{tab:longterm}).
With the same Monte Carlo, \cite{wass23} estimated a $\lambda_{NET}=29.3$ s$^{-1}$ and a $\lambda_{EFF}=390$ s$^{-1}$ for the proton flux in June 2017. In the same solar modulation condition, the proton flux simulated with our Toolkit returns a result of $\lambda_{NET}$=31.0 $\pm$ 1.0 s$^{-1}$ and $\lambda_{EFF}$ = 375 $\pm$ 14 s$^{-1}$ with an exceptional agreement despite the different spacecraft geometries considered in the simulations.
\begin{figure}
\centering
\includegraphics[width=0.5\textwidth]{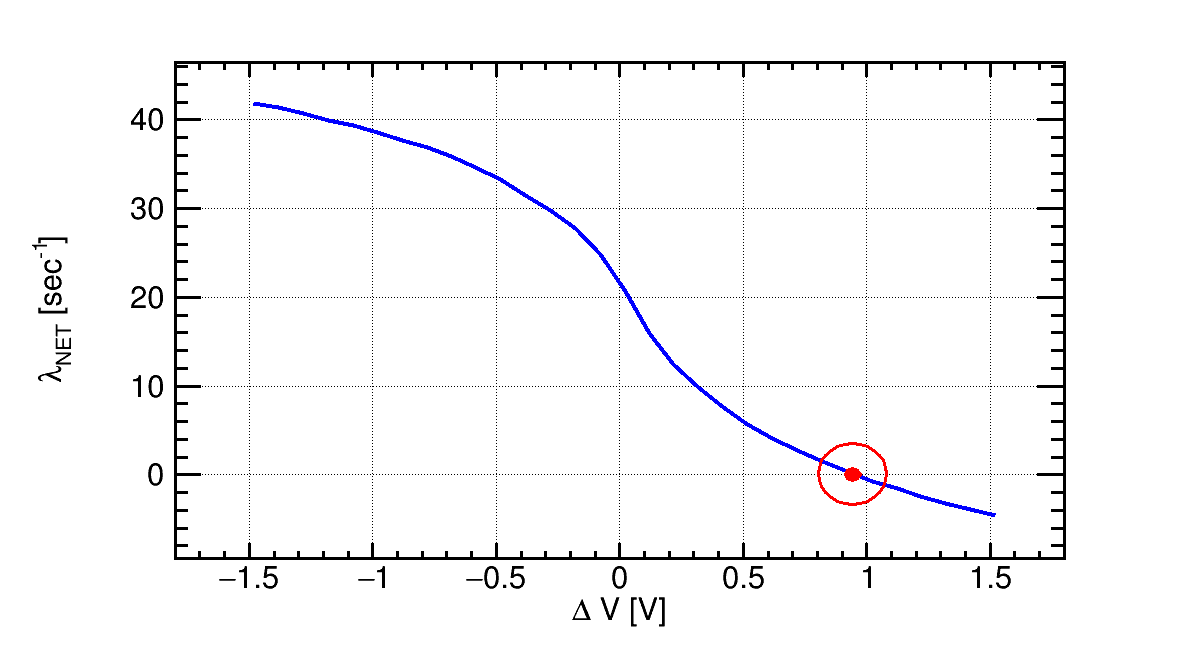 }
\caption{Simulation of the TM net charging evolution as a function of the TM to EH ground electric potential for a proton flux corresponding to the initial operation timeframe of LPF.}\label{fig:eq_pot}
\end{figure}
The consistency of the two GEANT4 Monte Carlo is also confirmed by the results of the simulation of the TM environmental charging evolution as a function of the TM potential shown in Figure \ref{fig:eq_pot}. The data points in the Figure were calculated using the Multiphysics software COMSOL propagating the low energy electrons in the electric field calculated for various TM potential. We recall that in LPF we noticed that the TM charge rate depended on the TM potential: the rate approached zero (TM equilibrium potential) for $V_{TM}\approx 0.9 $ V, very close to the value calculated by \cite{wass23} and that found by our simulation (see Figure \ref{fig:eq_pot}). We note that this was one of the main argument in favor of the presence of low energy electrons in the gap between the TM and the EH and for including in the old versions of the Monte Carlo the physics of the low energy electrons. 

Despite all this, our Monte Carlo simulation toolkit still needs improvements.
The comparison of the results reported in the Tables \ref{tab:longterm}, \ref{tab:SEP2}, and \ref{tab:resultLEI} shows that while the net charging results are similar for GEANT4 and FLUKA/LEI calculations, the effective charging in FLUKA/LEI appears, on average, higher by a factor of 2 and in a better agreement with the observations gathered on board LPF \citep{Armano2017,bridge}. It is worthwhile to stress that this discrepancy is much more evident with GCRs with respect to SEPs, due to a more relevant production of secondary electrons \citep[see][and Fig.\ref{fig:solp290989}]{bridge}. 

We have investigated the possible origin of the discrepancies between FLUKA/LEI and TMCTK.
Part of the difference in the results could be due to the spacecraft geometry (FLUKA/LEI uses a realistic reproduction of LPF geometry). Nevertheless, as stated before, we studied the dependence of TMCTK results on the material distribution around the TMs: we changed the shell material amount and/or their radius finding no relevant differences.
\begin{figure*}
\centering
\includegraphics[width=0.7\textwidth]{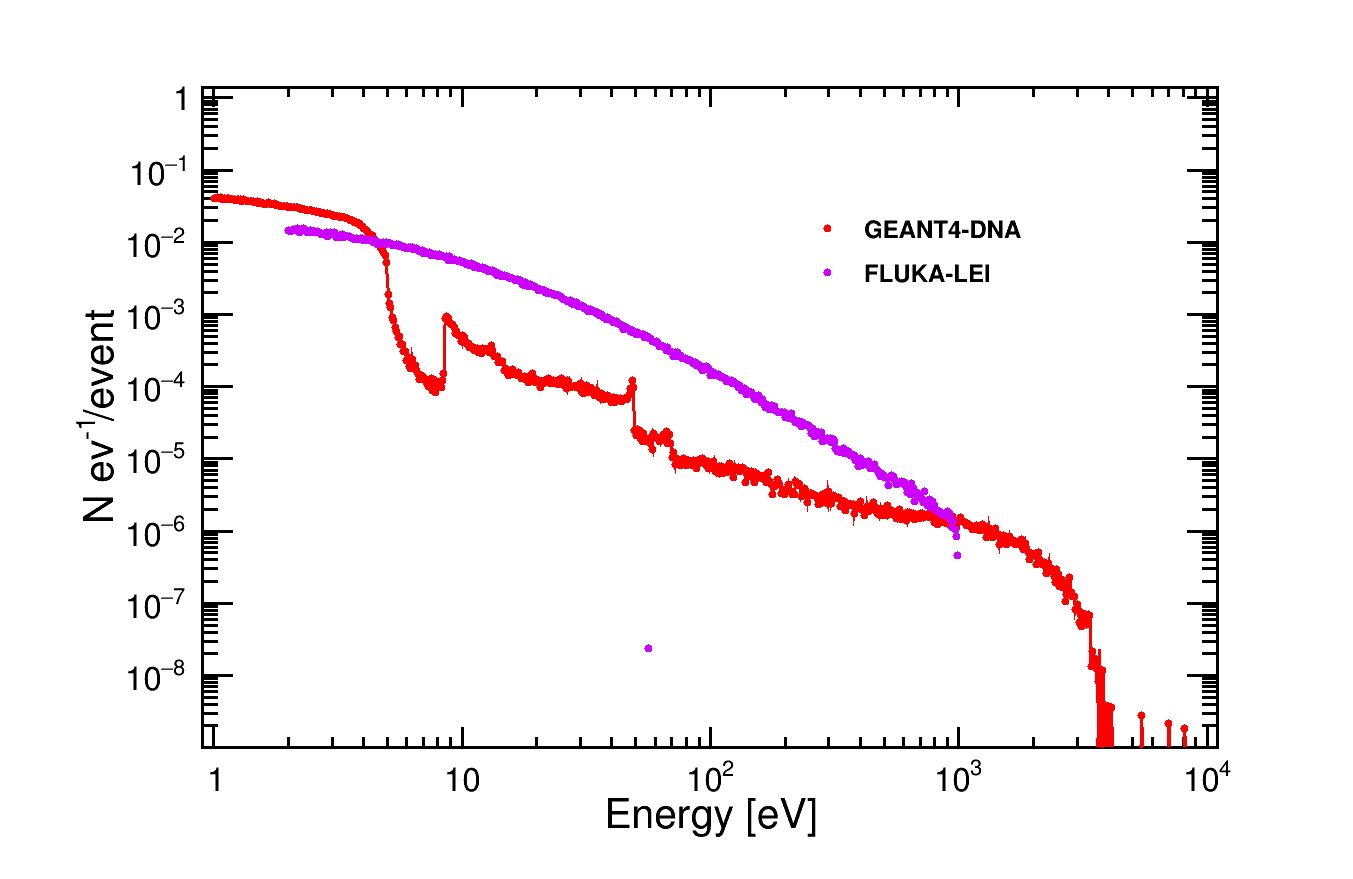}
\caption{Comparison between the energy spectra of secondary electrons per incident primary electron transmitted by a 150 nm thick gold target simulated using GEANT4-DNA (red curve) and LEI (purple curve).}\label{fig:spectrum_secondary}
\end{figure*}

\begin{figure}
\centering
\includegraphics[width=0.5\textwidth]{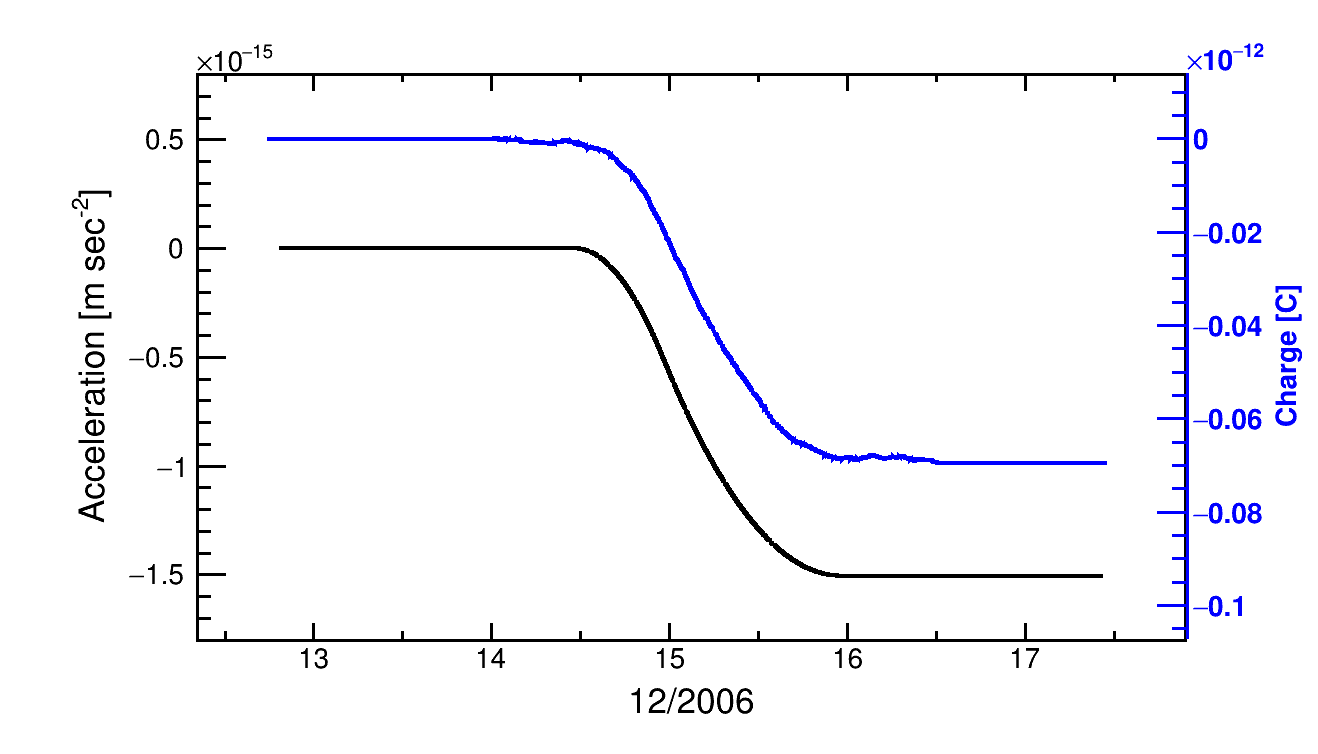 }
\caption{Forbush decrease contribution to the TM accumulated charge (blue line) and to TM acceleration (black line).}\label{fig:Forb_sig}
\end{figure}

\begin{figure}
\centering
\includegraphics[width=0.5\textwidth]{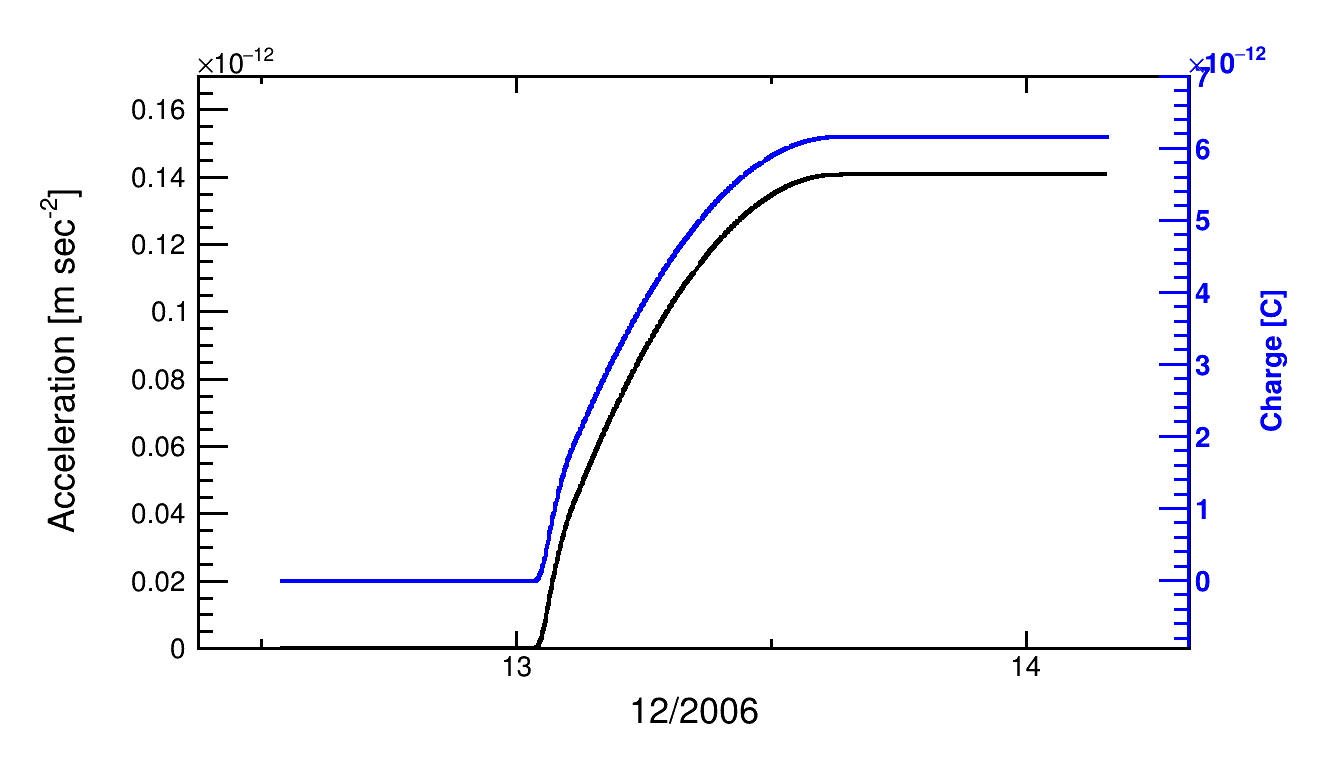 }
\caption{SEP contribution to the TM accumulated charge (blue line) and to TM acceleration (black line).}\label{fig:SEP_sig}
\end{figure}

The main difference between FLUKA/LEI and GEANT4-DNA is ascribable to a much larger  number of secondary electrons surrounding the test masses in FLUKA/LEI.
In the Introduction we stressed that the hadrons ionization in GEANT4 presents a sharp cut-off at the average ionization potential in gold of 790 eV, while in FLUKA/LEI the implementation is carried out down to 10 eV.
Moreover,  a simple experiment was carried out  to compare the performance of the two tools with primary electrons. We simulated a beam of 10 keV electrons crossing 100 nm thick gold slab.
In Figure \ref{fig:spectrum_secondary} we compare the spectrum  of secondary electrons per incident particle transmitted by the target: the spectrum associated with the GEANT4-DNA simulation exhibits a marked excess at a few eV and appears approximately one order of magnitude  below the LEI outcome, in particular below 1 keV. Moreover, the  GEANT4-DNA spectrum shows sharp step variations, typical of discontinuous parameterizations of cross sections and/or electron mean free path in the 10--100 eV energy interval.

\begin{figure}
\centering
\includegraphics[width=0.5\textwidth]{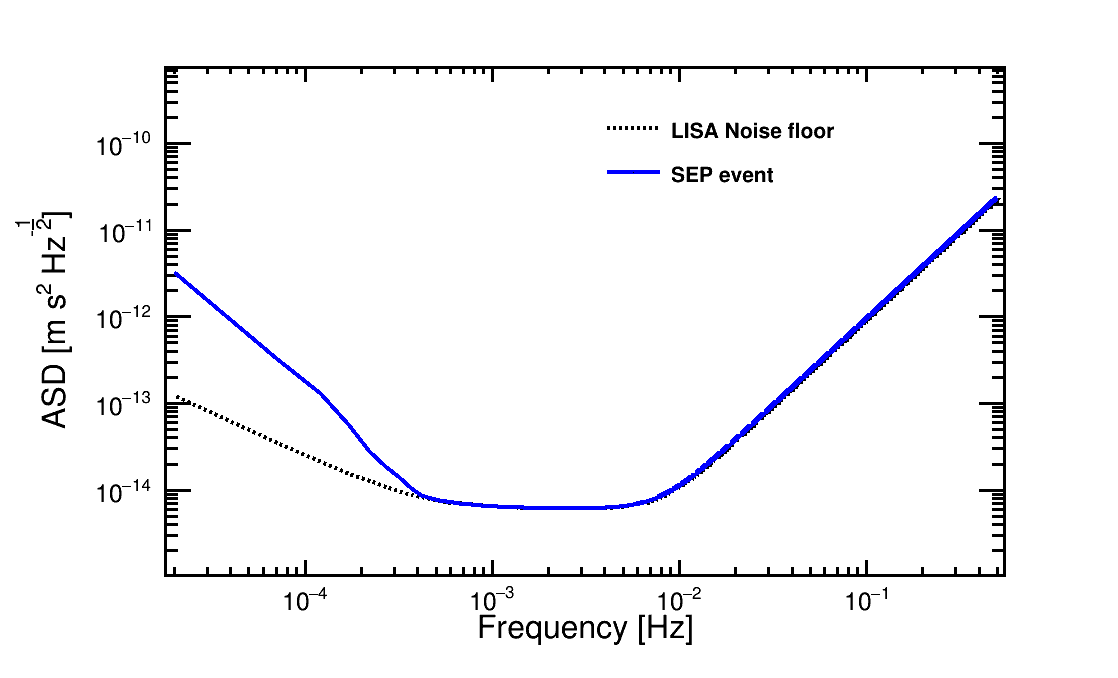 }
\caption{Amplitude spectral density of the LISA force noise and the SEP force signal in a time window of $2\times 10^5$ seconds.}\label{fig:LISA_asd}
\end{figure}

\section{Impact on LISA sensitivity}\label{sensitivitysection}

In the previous sections we have discussed the LISA TM charging  due to the action of particle fluxes of galactic and solar origin. We now  analyse the impact of the charging process on the LISA mission performance. The LISA TM is subjected to a Coulomb force due to the interaction of the TM charge with any non-zero average electrostatic field, from non-uniform surface "patch potentials" \citep{PhysRevLett.108.181101} and/or any applied fields. The formula for the force along the LISA sensitive axis, F$_x$, is:
 \begin{equation}
     F_x = Q_{TM}E_x= Q_{TM}\frac{\delta C_x}{\delta x}\frac{\Delta_x}{C_{TOT}}\,,
     \label{eqn:Fx}
 \end{equation}
where the electric field x-component E$_x$ is expressed in terms of the x-electrode derivative $\frac{\delta C_x}{\delta x}\approx 300\,$ pF/m, the TM self capacitance $C_{TOT}\approx 34\,$ pF, and the DC bias voltage $\Delta_x$ normalize the stray electrostatic field component along the $x$ axis to an electrostatic potential applied to a single $x$-face GRS electrode \citep{antonucci2011,PhysRevLett.108.181101}. A charge event occuring due to GCR flux short term variations or SEPs, makes $Q_{TM}$ changing with time producing a force signal. The time shape of the force is entangled to the time characteristics and to the energy content of the charge event. We have seen in Section \ref{studysection} that we are able to reconstruct the evolution of the charging signal (see for example Figure \ref{fig:lambdatimeevo}), by using Monte Carlo simulations of the different temporal phases of the event. From that, the force timeseries can be calculated according to Eqn. \ref{eqn:Fx}.
DC bias $\Delta_x\approx100$ mV has been observed in LISA-like GRS and electrostatically compensated to $\Delta_x=5$ mV level in laboratory and during LPF flight operations \citep{PhysRevLett.108.181101,armano2017b}. We assumed $\Delta_x=5$ mV, the expected value during LISA science operations.

Figures \ref{fig:Forb_sig} and \ref{fig:SEP_sig} show the evolution of the induced force on the LISA test masses for  the event of December, 13-15 2006, including SEP and FD.
The dynamics of the Forbush decrease was reconstructed as described in Sect. \ref{spaceenvsec} and shown in Figure \ref{NORECFD}. The flare  associated with the December 13, 2006 SEP event  was observed at 02:14 UT from the active region NOAA 10930. A  coronal mass ejections accelerating protons observed with PAMELA between 03:18 UT and 03:49 UT  was detected by LASCO at 02:54, thus revealing a very good magnetic connection between the active region and the PAMELA satellite. The time lag between onset, peak and decay phases of the SEP event are set on the basis of proton flux measurements carried out with PAMELA (see Figure \ref{fig:solp290989} and \citet{pamFD}).

For LISA performance, it is interesting to calculate the signal-to-noise ratio (SNR) of the charging events:
\begin{equation}
    (SNR)^{2} = \int_{f_{min}}^{\infty} \frac{\left|FFT\left(F_x\right)\right|^2}{S_n} df\,,
\end{equation}
with $f_{min}=20$ $\mathrm{\mu Hz}$ and 
$S_n \approx 4S_g\left(1+\left(\frac{0.4\mathrm{mHz}}{f}\right)^2\right)\left(1+\left(\frac{f}{8\mathrm{mHz}}\right)^2\right)$  (see Figure \ref{fig:LISA_asd}) expressing the low frequency limit of the LISA noise sensitivity for the X-Michelson Time Delay Interferometry \citep[TDI,][]{PhysRevD.59.102003,PhysRevD.107.082004} combination in terms of the residual force noise on a single TM\footnote{The factor $4$ multiplicating $S_g$ accounts for the contribution of the four TMs involved in the Michelson interferometer.}, where $S_g^{1/2}=3$ fm s$^{-2}$ Hz$^{-1/2}$ \citep{armano2017a,Nam,lisa_redbook}. At these frequencies the observatory sensitivity is completely dominated by TM acceleration noise, as so we neglect the interferometry noise contribution in this calculation \citep{Muratore}.
The SNR calculation yields 0.1 and 20 for the FD and the SEP event, respectively, indicating that LISA will be able to clearly observe the solar event. Were DC biases $\Delta_{x}$ larger than the $5$ mV limit because of lack of compensation, also Forbush decreases could pollute LISA data.

In that calculation, we have neglected the effect of the charge control system present in LISA, with the expectation that the response time of a possible continuous discharge will be slow enough, of order 10$^5$ seconds, to not have a large impact on "filtering" the SEP charge dynamics in the LISA band above 10$^{-4}$ Hz. As such, the SNR should not be much different in the case of continuous or intermittent discharge.

 We recall that the rate of SEP events with particle acceleration above tens of MeV is of the order of ten per year, at least for the transfer phase of the LISA satellites and the mission commissioning phase; the rate decreases to 8.5 at the beginning of the science data-taking period to about 1 at the end of the nominal LISA observation time \citep{grimani2025cqg}. The capability to discriminate these spurious events with respect to some gravitational ones lays on on-board measurements of the satellite radiation monitors and on the Monte Carlo simulations that would allow for reconstructing the event and inform the analysis of LISA data \citep{Nam}. Additionally it would be possible to continuously monitor the TM charge such as to allow possible subtraction of the force disturbance from the observatory TDI time series.

Ultimately, it is worth to note that from Eqn. \ref{eqn:Fx}, fluctuations on the TM charging process, here accounted with the $\lambda_{EFF}$ parameter, would generate force noise on the TM. Although our simulation have shown some limit in reproducing LPF measurements, we can use FLUKA/LEI results, as those shown in Tables \ref{tab:resultLEI} and \ref{tab:SEPLEI} to conclude that under GCR flux $\lambda_{EFF}$ won't exceed 1500 $s^{-1}$. This number appears close to what LPF has measured  (see Table \ref{tab:LPF_Results}) and would yield acceleration noise $S^{1/2}_{a} = \frac{\sqrt{2 e^2 \lambda_{EFF} }}{2\pi M_{TM}10^{-4}\mathrm{Hz}}\frac{\delta C_x}{\delta x}\frac{\Delta_x}{C_{TOT}}\approx 0.3$ fm s$^{-2}$  Hz$^{-1/2}$
confortably within $3$ fm s$^{-2}$  Hz$^{-1/2}$ allowed in LISA for stray electrostatic forces at $0.1$ mHz, the lower end of the LISA measurement bandwidth.  

In SEP events, together with the net charge $\lambda_{NET}$, also the charge fluctuation $\lambda_{EFF}$  rises largely, reaching values up to 6$\times$10$^3$ s$^{-1}$ for medium strong event such that on 13 Dec. 2006, or even 6$\times$10$^4$ s$^{-1}$ for the strong event on 29th Sept 1989 (see Table \ref{tab:SEP2} and \ref{tab:SEPLEI}). For SEP event of longer timescales (3-5 days), the analysis of the impact of such event in LISA becomes more complicated. Several aspects should be in particular taken into account: what discharge control algorithm will be in use, what ability we will have in flight to adjust its parameters,what will be the equilibrium potential at which the TM would rise. We leave this study for a dedicated paper.

\section{Conclusions}

This work presents a toolkit for the estimation of the effects of the space environment on the charging of LISA TM. The toolkit is based on GEANT4 package with GEANT4-DNA module for the generation and the propagation of the secondary electrons with energies below 100 eV in the outer 150 nm of the gold plated layers of the TM and EH. Low-energy electrons have been found to play a key role in the TM charging from the comparison of experimental data gathered with LPF and pre-launch Monte Carlo simulations. The adoption of the DNA module with its more accurate simulation of very low energy electrons, returned a better agreement with  LPF experimental data in line with other GEANT4 implementations of low-energy electromagnetic physics. Despite that, about a factor of two mismatch remains in the estimate of the charging noise (and thus of $\lambda_{EFF}$) with respect to the LPF data. The good agreement between the outcomes of the FLUKA/LEI Monte Carlo and the LPF data suggests that DNA module presents a still incomplete description of the role of LEE in the TM charging process. This is confirmed from the comparison of FLUKA/LEI and GEANT4-DNA of the energy spectra of simulated electrons emitted from a 100 nm gold layer with incident electrons beams of 10 keV energy. This issue was acknowledged by scientists of the GEANT-4 collaboration and improvements are expected in the future release of the DNA module for low-energy electromagnetic physics in gold. It is important to stress that with respect to the TM 
net charging ($\lambda_{NET}$) estimates, GEANT4  and FLUKA/LEI Monte Carlo appear consistent  with each other and in  good agreement with LPF measurements.

We have used the toolkit to consider different case studies associated with the impact of the environment on the LISA mission. In particular, we have evaluated the TM charging due to GCRs under extreme solar modulation conditions, during recurrent and non-recurrent short-term variations of the GCR flux and during SEP events of different intensity. 
By extrapolating the acceleration noise budget of LPF TM to LISA, the TM charging due to GCRs for the extreme cases of solar activity is well below the 3 fm s$^{-2}$ Hz$^{-1/2}$ total acceleration noise requirement. 
LISA will be a signal dominated mission, short-term variations of incident particle fluxes may generate spurious signals mimicking genuine GWs. Whereas GCR recurrent short-term variations, having typical durations of 9.1 days, are completely out of the LISA sensitivity band, the sharp decrease observed during FDs (see Figure \ref{fig:Forb_sig}) generates a step-like signal that would appear in the LISA band. This signal could have SNR$>$1 with large stray electrostatic fields, $\Delta_{x}\geq$ 50 mV, within the few mm gap between EH and TM. 

SEP events have shown to produce increases of several orders of magnitude on net charging and charge noise over typical time scales from a few hours to a five days depending on the magnetic connection of the spacecraft to an active region of the Sun from which several events are generated in a row. We have found that a medium-strong event of duration of a few hours, such as that measured by PAMELA in Dec. 2006, could generate a strong SNR$\approx$20 signal in the LISA band, which could require vetoing the relevant portions of the datastream.

In addition to spurious signal generation, SEP events with longer time scales may charge the TM to potentials as large as 
1 V affecting  the mission operation. Therefore, for LISA it will be important to analyse also the response of the charge management system to these events. 


\begin{acknowledgements}
The authors thank E. Castelli (NASA) for helpfull discussions, suggestions and simulations regarding the impact of the charging signal to the LISA sensitivity. This work was funded under the European Space Agency contract No. 4000133571/20/NL/CRS - TEST MASS CHARGING TOOLKIT AND LPF LESSONS LEARNED. F. Dimiccoli and V. Ferroni were funded by ASI - Agenzia Spaziale Italiana - ACCORDO ATTUATIVO n. 2024-36-HH.0
dell’ACCORDO QUADRO ASI/Università di Trento n. 2017-32-H.0 - Addendum n.2017-29-H.1-2020 All’Accordo n. 2017-29-H.0

\end{acknowledgements}

%
%
  \bibliographystyle{aa} 
  \bibliography{aa} 



\end{document}